\documentclass[12pt]{iopart}

\usepackage{todonotes}

\expandafter\let\csname equation*\endcsname\relax

\expandafter\let\csname endequation*\endcsname\relax

\usepackage{amsmath}
\RequirePackage{amssymb}
\usepackage{textcomp}

\usepackage{graphicx}
\usepackage{bm}
\usepackage{upgreek}
\DeclareSymbolFont{matha}{OML}{txmi}{m}{it}
\DeclareMathSymbol{\varv}{\mathord}{matha}{118}

\newcommand{\figref}[1]{Fig.~\ref{#1}}

\definecolor{grayblue}{rgb}{0.85 , 0.9, 1}

\newcommand{\quotes}[1]{``#1''}
\newcommand{\deltaV}{{\ensuremath{\triangle V}}}
\newcommand{\deltapsi}{{\ensuremath{\triangle \psi}}}
\newcommand{\psimicro}{{\ensuremath{\psi^{\rm{micro}}}}}

\newcommand{\Br}{{\boldsymbol{r}}}

\newcommand{\totdiff}{\textrm{d}}

\newcommand{\ds     }{\ensuremath{\totdiff{s}}}

\newcommand{\dx     }{\ensuremath{\totdiff{x}}}
\newcommand{\dy}{\ensuremath{\mathrm{d}y}}
\newcommand{\dz}{\ensuremath{\mathrm{d}z}}

\newcommand{\gndrhobold}{\boldsymbol{\varrho}}
\newcommand{\gndrho}{\varrho}

\begin{document}

\title{Data-mining of dislocation microstructures: concepts for coarse-graining of internal energies}

\author{Hengxu Song$^{1, 4}$, Nina Gunkelmann$^{1, 2}$, Giacomo Po$^{3}$, and Stefan Sandfeld$^{1,4*}$}

\address{$^{1}$ The Micromechanical Materials Modelling Group, Institute of Mechanics and Fluid Dynamics, TU Bergakademie Freiberg, 09599 Freiberg, Germany}
\address{$^{2}$ Institute of Applied Mechanics, Computational Material Sciences/Engineering, Clausthal University of Technology, 38678 Clausthal-Zellerfeld, Germany}
\address{$^{3}$ Department of Mechanical and Aerospace Engineering, University of Miami.}
\address{$^{4}$ Institute for Advanced Simulations -- Materials Data Science and Informatics (IAS-9), Forschungszentrum J{\"u}lich GmbH, 52425 J{\"u}lich, Germany}
\ead{s.sandfeld@fz-juelich.de}

\vspace{10pt}

\begin{abstract}
	Continuum models of dislocation plasticity require constitutive closure assumptions, e.g., by relating details of the dislocation microstructure to energy densities. Currently, there is no systematic way for deriving or extracting such information from reference simulations, such as \emph{discrete dislocation dynamics} or \emph{molecular dynamics}.	
	Here, a novel data-mining approach is proposed through which energy density data from systems of discrete dislocations can be extracted. Our approach relies on a systematic and controlled coarse-graining process and thereby is consistent with the  length scale of interest. 
	For data-mining, a range of different dislocation microstructures that were generated from 2D and 3D discrete dislocation dynamics simulations, are used. The analyses of the data sets result in energy density formulations as function of various dislocation density fields. 
	The proposed approach solves the long-standing problem of voxel-size dependent energy calculation during coarse graining of dislocation microstructures. Thus, it is crucial for any \emph{continuum dislocation dynamics} simulation. 
\end{abstract}
%

\section{Introduction}
Plasticity in crystalline materials originates from the collective movement of dislocations. Understanding how dislocations evolve as a system is the key for many pratical problems, such as material strength and failure. However, many of the details of the evolution of systems of dislocations are mostly inaccessible to experimental approaches due to the very small length -- and sometimes also time -- scale of dislocation evolution. Therefore, computational studies of dislocation plasticity can be the key for advances in material science and engineering. Both 2D~\cite{vandergiessen1995,cleveringa1999discrete, papanikolaou2017obstacles, song2019universality} and 3D Discrete Dislocation Dynamics (DDD) methods~\cite{bulatov2006computer, cui2016controlling,Stricker:2015aa, Fivel2010} have been used to study plasticity especially at small length scales and provide many fundamental understandings of the materials. However, due to the fact that each single dislocation is resolved individually, the computational cost of DDD scales almost quadratically with the number of dislocations (discretized dislocation segment in 3D) under consideration. Therefore, when it comes to a large number of interacting dislocations (high dislocation densities/large simulation volume), the method is computationally very expensive even with the most advanced computational technique~\cite{bertin2019gpu}. At the same time, in particularly the situations with denser dislocation structures strongly suggest that continuum approaches based on statistical theories of dislocations or dislocation dynamics might be beneficial. 
From a computational perspective, reducing dislocation dynamics to a system of partial differential equations amenable to solution by standard continuum computational methods would be highly desirable.

There has been significant effort towards studying dynamics of dislocation densities and developing ``continuum dislocation dynamics" theories and simplified models. Based on systems of straight parallel edge dislocations (i.e., 2D systems of dislocations), Groma was the first to derive the evolution equations of dislocation density~\cite{groma1997link} and later on obtained pair correlation functions based on the statistical evaluation of a large number of discrete dislocation simulations~\cite{groma2003spatial} whose analysis resulted in terms governing the interaction of systems of dislocations. 
These 2D studies have been quite successful and hinted on potential developments of similar approaches in 3D. Unfortunately, the direct extension of approaches from 2D to 3D is practically still impossible. This difficulty has its roots in the definition of a suitable, three-dimensional continuous dislocation measure as well as in the huge technical difficulty of obtaining the correlation functions in 3D. The former is due to the fact that unlike in 2D, it is difficult in 3D to define simple density variables to reflect the kinematics of moving curved dislocation lines. Only recently, the so-called continuum dislocation dynamics (CDD) proposed by Hochrainer and coworkers~\cite{Hochrainer2007, Hochrainer2014, Sandfeld2011} has introduced dislocation density and density-like variables along with their evolution equations which advanced the development of a general continuum description of 3D curved dislocations. The shortcoming of this theory up to now is that no systematic approach for relating the averaged dislocation microstructure to the velocity, by which dislocation densities move, exists. Therefore, providing this missing link is critical to fulfill the ``dynamic closure''. The recent progress made by Hochrainer and Zaiser~\cite{hochrainer2016thermodynamically,zaiser2015local} has elucidated the theoretical basis of the derivation of dislocation driving forces utilizing the free energy of the system.
In the case of isothermal elasticity, the free energy density is essentially the strain energy density of the system. However, it is not trivial to quantify the strain energy density of a coarse grained dislocation system due to the ``mesh-dependency", i.e., the results strongly depend on the chosen voxel size during the coarse graining.  The problem has its roots in the nature of the coarse graining process: microstructure information below the average voxel size is almost completely lost. Therefore, the first step towards the fulfillment of the ``dynamic closure'' is to obtain a ``mesh-independent" strain energy density, i.e., recover the missing energy density during coarse graining.

In this work, a new data-mining approach is introduced which extracts energy density data from systems of discrete dislocations. Once the functional form of the energy density as a function of the CDD field variables is established, stresses can be obtained which then would result in velocities. This approach is consistent with the coarse-graining of the geometrical properties (i.e., the different density fields). Our analysis is based both on numerically generated 2D and 3D dislocation microstructures. The dislocation microstructures are realistic in the sense that they are generated through benchmarked discrete dislocation dynamics simulations. The 2D dislocation microstructures resulted from both uniaxial tension and pure bending simulations. The 3D dislocation microstructures encompass random loop structures (without relaxation) and relaxed dipolar loop structures. Utilizing these dislocation structures in a data-driven approach, we relate the strain energy density of the system to the coarse grained CDD field variables through a functional form of the CDD field variables. Our approach solves the long-standing problem of the ``mesh-dependency" during the coarse graining of dislocation microstructures: given the desired voxel size, our results are able to recover the lost, mesh-dependent strain energy density during the coarse graining, thus provide the accurate strain energy density that is effectively independent of the chosen voxel size. This is an important prerequisite for the ``dynamic closure'' of a continuum dislocation dynamics framework.

This paper is structured as follows: in section 2, we provide an overview of the CDD theory and problem formulation. In section 3, we firstly introduce the problem of losing important information during coarse graining in the context of obtaining the strain energy density in a coarse grained system, then we explain in detail our data-mining strategy to relate the system strain energy density to the CDD field variables. Section 4 shows an application of the proposed data-mining strategy on 2D discrete dislocation systems, furthermore, the finite boundary effect of the system is also discussed. In section 5, we show the feasibility of the proposed strategy for 3D discrete dislocation systems.  In section 6, energy formulations considering geometrically necessary dislocations (GND) density and dislocation curvature density are examined. In section 7, we summarize and discuss the results. The appendix includes more details on finite boundary effect of the dislocation system and the choice of the reference voxel size during the data-mining process.

\section{Theoretical background: overview of the CDD theory and problem formulation.}
The CDD theory essentially represents a discrete dislocation system as a set of continuous, density-like field variables, the temporal change of which is governed by a set of evolution equations. One of the simplest descriptions includes the total dislocation density $\rho$, the curvature density $q$, and a vector of \quotes{geometrically necessary} signed densities (GND density) $\gndrhobold$ which have the average line orientation $\bm{l}_{\gndrhobold}=\gndrhobold/ \gndrho$ with $\gndrho=|\gndrhobold|$. The reference coordinate system may be oriented such that the components of $\gndrhobold$=[$\gndrho_{s}$, $\gndrho_{e}$] represent the orientation of excess screw and edge dislocations.
The temporal evolution of these field quantities are -- in local slip system coordinates -- governed by the following set of equations:
\begin{equation}
{\partial_{t} \rho} = -\nabla \cdot (\varv \gndrhobold^{\perp}) + \varv q
\end{equation}
\begin{equation}
{\partial_{t} \gndrhobold} = -\nabla \times (\varv \rho \boldsymbol{n})
\end{equation}
\begin{equation}
{\partial_{t} q} = -\nabla \cdot (\varv \boldsymbol{Q}^{(1)}+\boldsymbol{A}^{(2)}\cdot \nabla \varv)
\end{equation}
{where $\boldsymbol{n}$ is the slip plane normal.  The vector $\gndrhobold^{\perp}=[\gndrho_{\rm{e}}, -\gndrho_{\rm{s}}]$ is the GND density vector rotated by 90$^{\circ}$. $\boldsymbol{Q}^{(1)}$ is assumed to be equal to -$\gndrhobold^{\perp}q/\rho$ while $\boldsymbol{A}^{(2)}$ can be obtained from a `maximum entropy' approach $\boldsymbol{A}^{(2)}=\frac{\rho}{2} \left[ (1+\Phi)\boldsymbol{l}_{\gndrho} \otimes \boldsymbol{l}_{\gndrho}  + (1-\Phi) \boldsymbol{l}_{\gndrho^{\perp}} \otimes \boldsymbol{l}_{\gndrho^{\perp}} \right]$ where $\boldsymbol{l}_{\gndrho^{\perp}}$ is the unit vector perpendicular to $\boldsymbol{l}_{\gndrho}$ and $\Phi \approx (\gndrho/\rho)^{2}(1+(\gndrho/\rho)^4)/2$. For more details, the readers can refer to~\cite{sandfeld2015microstructural} and references therein.}

This set of equations is able to predict the evolution of density-microstructure if, and only if, the velocity $\varv$ is given. {The ability to evolve dislocations as curved and connected lines with the concomitant line length change during motion in a given velocity field is required to satisfy the criterion of \textit{kinematic consistency}.} The \textit{kinematic consistency} of CDD has been recently verified by Sandfeld and Po~\cite{sandfeld2015microstructural} through a direct comparison of dislocation density field evolution under idealized velocity fields $\varv$ between CDD and 3D DDD without considering dislocation interactions. 
However, the CDD theory itself does not say anything about \emph{how} the velocity is related to the microstructural variables. This issue is related to the other criterion that a CDD framework should satisfy, which is the  \textit{dynamic consistency}: a CDD framework should be able to predict the actual evolution of a dislocation system under externally applied mechanical load together with considering the mutual interactions between dislocations. This is crucial because it requires to take into account the effects of short range interaction stresses which act at a scale below the averaging size of the continuum representation.
The dynamic consistency essentially requires the derivation of dislocation driving forces $\uptau_{\rm{net}}$, which can be used to obtain $\varv$ from the generic form 
\begin{equation}
\varv = M (\uptau_{\rm{net}}, \rho ) \, b\, \uptau_{\rm{net}}
\end{equation}
with $M (\uptau_{\rm{net}}, \rho )$ being the dislocation mobility function.  A recent approach based a thermodynamic ansatz in~\cite{hochrainer2016thermodynamically}  makes it possible to obtain the driving force as a function of the derivatives of the free energy density $\psi$ with respect to the  CDD field quantities 
\begin{equation}
\uptau_{\rm{net}} = \hat f\left(\rho, \gndrhobold, q; \frac{\partial \psi} {\partial \rho}, \frac{\partial \psi} {\partial \gndrhobold}, \frac{\partial \psi} {\partial q} \right)
\end{equation}
where $\hat f$ is a function depending on the given variables and derivatives. Therefore, the detail of the CDD theory that is missing for the dynamic closure is how to express a system's strain energy density in terms of its coarse grained field quantities, the density and density-like variables. In particular, the system's strain energy density has to account for the information that has been lost during the coarse graining process, i.e., it should -- in numerical terms -- become ``mesh-independent''.

\section{Data-mining strategy}
\subsection{General idea}
The free energy plays a crucial role in most continuum plasticity frameworks. For example, in crystal plasticity, Gurtin~\cite{gurtin2008finite} formulated the total free energy density
$\psi$ as the standard elastic strain energy density augmented by a defect energy density
\begin{equation}\label{typical-energy-density}
\psi = \psi^{\rm{el}} + \psi^{\rm{defect}},
\end{equation}
where both $\psi^{\rm{el}}$ and $\psi^{\rm{defect}}$ are function of GND density only. Kooiman et al.~\cite{kooiman2015microscopically} derived an explicit expression for the free energy of straight and parallel dislocations with different Burgers vectors, and they found that the defect energy density depends on both positive and negative dislocation density independently. Above mentioned works are theoretical studies of the framework, therefore did not involve the discussion of an unavoidable technical issue when one implements the framework numerically: coarse graining. Density quantities in those frameworks are coarse grained quantities by definition. In the CDD framework, the CDD field variables are coarse grained quantities representing the dislocation microstructure based on averages within coarse graining sub-volumes -- which we will call \emph{voxels}, subsequently. The problem of coarse graining is that microstructural information below the coarse graining voxel size $\deltaV$ is missing. 
\begin{figure}[ht!] 
	\centering
	\includegraphics[scale=1.0]{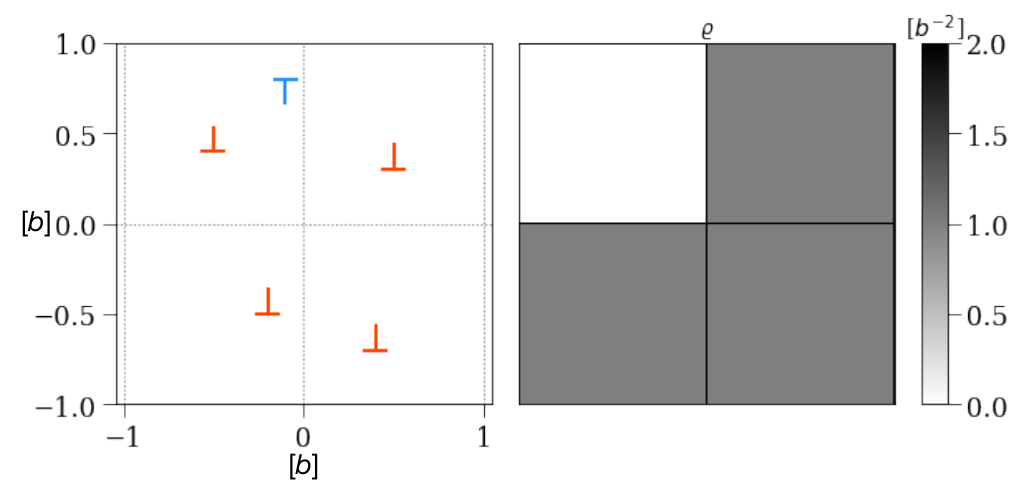}
	\caption{
        Coarse graining a dislocation microstructure (left) with averaging voxels of size $b$ indicated by the dashed lines results in the GND density shown on the right. Such coarse graining would completely lose the information of the dislocation dipole on the top left voxel and the difference of locations for single dislocations inside the other voxels. Any energy measure based on the GND density is afflicted by this information loss.}
\label{fig:figure1-2d-example}
\end{figure} 
For example, in \figref{fig:figure1-2d-example} with the squares being the coarse graining voxels, the information of a dislocation dipole will be completely lost since the GND density in the voxel is zero. For non-zero GND density, the information of the exact location of the dislocation is missing since they yield the same GND density after the coarse graining. It is straightforward to imagine that one would get different coarse graining dislocation densities if using different voxel size $\deltaV$, for example, using one voxel for the whole dislocation structure. Therefore, with a user-predefined voxel size $\deltaV$ in a numerical simulation, the energy density information should be written as 
$\psi_{\triangle V}$, $\psi^{\rm{el}}_{\triangle V}$, $\psi^{\rm{defect}}_{\triangle V}$. The direct consequence of the voxel size dependency is that different choices of the voxel size $\deltaV$ can result in different $\psi$ of the system which would yield different dynamics of the dislocation system. For a dislocation system, however, there should exist a unique free energy density that includes all microscale information of the system, and such free energy density can be numerically obtained when the coarse grained voxel size is small enough. For example, for a dislocation loop in Fig.~\ref{fig:3D-single-loop}, when the voxel edge length is as small as the magnitude of the Burgers vector and the effect of the dislocation core region (approximately indicated by 2. in Fig.~\ref{fig:3D-single-loop}) is ignored, the elastic fields of the dislocation loop are described in a nearly perfect manner and therefore coarse graining with these very fine ``reference voxels'' (indicated by inset 1. in Fig.~\ref{fig:3D-single-loop}) would not lose any relevant information of the dislocation. However, using these ``reference voxels'' for a complex dislocation microstructure in a large simulation domain is not computationally feasible, there one would prefer a much larger coarse graining voxel as indicated by 3. in Fig.~\ref{fig:3D-single-loop}. With the large voxel size $\Delta V$, following Gurtin~\cite{gurtin2008finite}, one can then calculate $\psi^{\rm{el}}_{\triangle V}$ and $\psi^{\rm{defect}}_{\triangle V}$ which eventually end up with a voxel size dependent $\psi_{\triangle V}$.   
\begin{figure}[ht!] 
	\centering
	\includegraphics[scale=0.8]{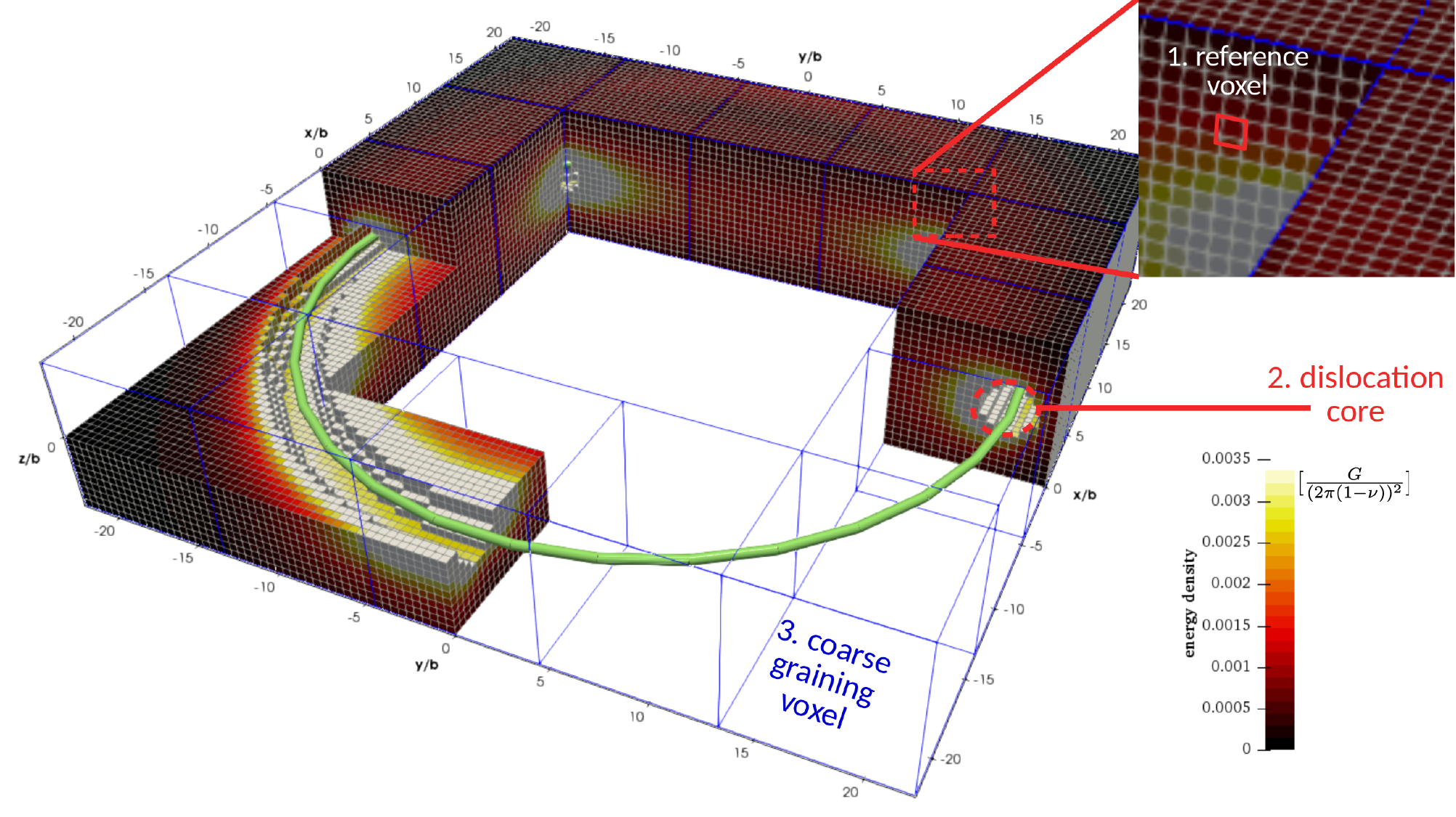}
	\caption{ Sketch of the averaging procedure: the energy density around a dislocation is sampled with a high resolution ($\approx$ 1$b$) as indicated by the small, colored voxels (1.). Only the core region (2.) of the dislocation is excluded (note the missing voxels in the vicinity of the dislocation line). Averaging the energy data inside the coarse graining volumes (the larger volumes drawn in blue, (3.))}
\label{fig:3D-single-loop}
\end{figure} 
Here, we reinterpret the problem in a different manner: the defect energy density $\psi^{\rm{defect}}$ can be understood as the difference between the free energy density and the elastic strain energy density. We know that numerically both $\psi^{\rm{el}}$ and $\psi^{\rm{defect}}$ are voxel size dependent in the sense that how much these quantities contain the information of the real physical system strongly depend on the voxel size. However, the sum of the two, i.e., $\psi$ should be an objective unique quantity of the system which contain all microscale information of the system. Accurate $\psi$ of a dislocation microstructure can be numerically resolved only by using very small voxel size, as shown for a dislocation loop in  Fig.~\ref{fig:3D-single-loop}, while what is actually needed in a simulation with voxel size $\Delta V$ is the $\psi$ for $\Delta V$ which we term as $\psimicro_{\Delta V}$ which stands for the free energy density on the length scale of $\Delta V$ that includes the microstuctural information below $\Delta V$. 
Here, we approximate $\psi^{\rm micro}_{\Delta V}$ by averaging all $\psi_{\Delta V = b^{3}}$ in $\Delta V$, as shown in Fig.~\ref{fig:3D-single-loop} (averaging all reference voxel data in a coarse graining voxel). Therefore we can rewrite \eqref{typical-energy-density} for a numerical implementation as 
\begin{equation}\label{energy-equation}
\psi^{\rm{micro}}_{\triangle V} = \psi^{\rm{meso}}_{\triangle V} + \deltapsi.
\end{equation}
Here we rewrite $\psi^{\rm{defect}}_{\triangle V}$ as $\deltapsi$ because now it can be numerically solved through $\psi^{\rm{micro}}_{\triangle V}-  \psi^{\rm{meso}}_{\triangle V}$. Unlike Gurtin~\cite{gurtin2008finite} who intepreted the defect energy density as the function of only GND densities, the defect energy density in our interpretation takes the same energy functional as the free energy of a dislocation system~\cite{zaiser2015local, Wilkens1969, berdichevsky2006continuum, groma2016dislocation} which is the function of the coarse grained densities, such as dislocation density, GND density and dislocation curvature density. With the numerical value of $\deltapsi$ and specific given energy functional, we can then obtain the coefficient (weight factor) of $\deltapsi$ energy functional.  Such process can be repeated for a variety of $\Delta V$, i.e., we can eventually obtain the dependence of coefficient (weight factor) of $\deltapsi$ energy functional on $\Delta V$. This result is crucial because after we have such dependence, for a dislocation microstructure with the computationally desired voxel size $\Delta V$, we can then calculate $\psi^{\rm{meso}}_{\triangle V}$ following the classical continuum theory of dislocations and $\psi^{\rm{defect}}_{\triangle V}$ using the obtained coefficient (weight factor) corresponding to $\Delta V$, we can then obtain $\psi^{\rm{micro}}_{\triangle V}$ which includes the micro-structure information below the averaging voxel size and therefore guarantees a correct dynamics of the coarse grained dislocation system. 

\subsection{Detailed explanation of the general data-mining strategy}

For a given 3D dislocation structure, a mathematical density measure along the dislocation line can be defined as 
\begin{equation}\label{density-measure}
\delta_{\bm{c}}(\bm{r}) = \int\limits_{0}^{L_c}\delta(\bm{r}-\bm{c} (s))\,\ds.
\end{equation}
Here $\bm{r}$ is a point in the 3D space, $L_{c}$ is the total length of dislocation curve $\bm{c}(s)$ and $\delta()$  is the three-dimensional Dirac delta function. Using this measure, we may write the discrete (indicated by superscript `d') dislocation density as 
\begin{equation}\label{discrete-dislocation-density}
\rho^{\rm{d}}=\sum_{\boldsymbol{c}} \delta_{\bm{c}}
\end{equation}
and discrete GND density as 
\begin{equation}\label{discrete-GND-density}
\gndrhobold^{\rm{d}}= \sum_{\boldsymbol{c}} \delta_{\bm{c}} \frac{\mathrm{d}\bm{c}} {\ds}, 
\end{equation}
where the sum runs over all dislocation lines $\bm{c}(s)$ in the system,  $\frac{\mathrm{d}\bm{c}} {\ds}$ is the local line orientation. Furthermore, we can also write discrete Kr\"{o}ner-Nye tensor as 
\begin{equation}\label{discrete-Nye-tensor}
\bm{\alpha}^{\rm{d}}= \sum_{\boldsymbol{c}} \delta_{\bm{c}} \frac{\mathrm{d} {\bm{c}}} {\ds} \otimes {\bm{b}}. 
\end{equation}

When the coarse graining is carried out for the dislocation system, we calculate the spatial average of density fields over some small averaging volume. We assume, without loss of generality, that the averaging volumes are rectangular cuboids (or \emph{voxels}) where the centers $\Br_{i}$ have the coordinate ($x_{i},y_{i},z_{i}$). Each voxel defines a domain $\Omega_{i}$: 
\begin{equation}
\Omega_{i}=
\left[x_{i}-\frac{\Delta x}{2} \ldots x_{i}+\frac{\Delta x}{2}\right]
\times 
\left[y_{i}-\frac{\Delta y}{2} \ldots y_{i}+\frac{\Delta y}{2}\right]
\times
\left[z_{i}-\frac{\Delta z}{2} \ldots z_{i}+\frac{\Delta z}{2}\right]
\end{equation}
with its volume being $V_{\Omega_i}=\Delta x\,\Delta y\,\Delta z = \Delta {V}$.
The continuous density fields can be then transformed into ``voxel data'' by applying the averaging operator in $\Omega_{i}$ which is defined as 
\begin{equation}
\langle\bullet\rangle_{\Omega_{i}} = \frac{1}{\deltaV}
    \int\limits_{x_{i}-\frac{\Delta x}{2}}^{x_{i}+\frac{\Delta x}{2}} 
    \int\limits_{y_{i}-\frac{\Delta y}{2}}^{y_{i}+\frac{\Delta y}{2}} 
    \int\limits_{z_{i}-\frac{\Delta z}{2}}^{z_{i}+\frac{\Delta z}{2}} \bullet \;\; \dx' \dy' \dz'.
\end{equation}
Within $\Omega_{i}$, the dislocation density can then be calculated as
\begin{equation}\label{dislocation-density}
\rho=\left< \rho^{\rm{d}}  \right> = \left< \sum_{\boldsymbol{c}} \delta_{\bm{c}} \right>
\end{equation}
Note that any coarse grained information in $\Omega_{i}$ is ``stored'' at the voxel center $\bm{r}_{i}$, i.e., we are using a constant approximation inside the averaging volume. Therefore the coarse grained dislocation density should be expressed as $\rho(\bm{r}_{i})$ but we drop $(\bm{r}_{i})$ for convenience and also to avoid possible confusion with the variable $\bm{r}$ in \eqref{density-measure}.  Similarly, the coarse grained GND density vector is 
\begin{equation}\label{GND-density}
\gndrhobold=\left< \gndrhobold^{\rm{d}}  \right> =\left< \sum_{\boldsymbol{c}} \delta_{\bm{c}} \frac{\mathrm{d}\bm{c}} {\ds} \right>
\end{equation}
and coarse grained Kr\"{o}ner-Nye tensor is 
\begin{equation}
\bm{\alpha}= \left< \bm{\alpha}^{\rm{d}}  \right> =  \left< \sum_{\boldsymbol{c}} \delta_{\bm{c}} \frac{\mathrm{d} {\bm{c}}} {\ds} \otimes {\bm{b}} \right>
\end{equation}
The GND density $\gndrho$ is its norm $\left\Vert \gndrhobold \right\Vert$. 
If all dislocations in $\Omega_{i}$ have the same Burgers vector $\bm{b}$, then $\bm{\alpha}$ can be written as 
\begin{equation}\label{Nye-tensor}
\bm{\alpha}=\gndrho\bm{l}\otimes\bm{b} 
\end{equation}
where the average line direction of dislocations in $\Omega_{i}$ is then $\bm{l}=\gndrhobold/ \gndrho$.
With the Kr\"{o}ner-Nye tensor, one can calculate the elastic distortion due to geometrically necessary dislocations (GNDs) through the Mura-Willis equation~\cite{mura2013micromechanics} and obtain the stress tensor:
\begin{equation}\label{stress-tensor}  
\sigma_{ij}(\bm{r})=C_{ijkl}\int \epsilon_{lnh} C_{pqmn}G_{kp,q}(\bm{r}-\bm{r^{\prime}})\alpha_{hm}(\bm{r^{\prime}})\mathrm{d} \bm{r^{\prime}} 
\end{equation} 
where  $\epsilon_{lnh}$ is the Levi-Civita symbol and the $\alpha_{hm}$ are the components of the Kr{\"o}ner-Nye tensor. The partial derivative of the Green's function $G$ has the following form: 
\begin{equation}
G_{kp,q}(\bm{r})=\frac{1}{16\pi\mu(1-\nu)}[2(1-\nu)\delta_{kp}\partial_{q}\Delta-\partial_{k}\partial_{p}\partial_{q}]\bm{r}
\end{equation}
where $\Delta$ is the Laplace operator. We employ isotropic elasticity, such that we can write the components of the stiffness tensor in terms of the Lam\'e parameters $\lambda$ and $\mu$: 
\begin{equation}
C_{pqmn}=\lambda\delta_{pq}\delta_{mn}+\mu(\delta_{pm}\delta_{qn}+\delta_{pn}\delta_{qm}).
\end{equation}
For each voxel one can then get the strain energy density:
\begin{equation}\label{energy-density}
\psi=\frac{1}{2}S_{ijkl}\sigma_{ij}\sigma_{kl},
\end{equation}
here ${S}_{ijkl} = {C}_{ijkl}^{-1}$ are the components of the elastic compliance tensor. 
Due to the shape of \eqref{energy-density} $\psi$ is effectively quadratic also in $\alpha_{ij}$. The consequence is that in general for the average strain energy it is
\begin{equation}
\left\langle \psi(\alpha_{ij})  \right\rangle \ne \psi(\left\langle \alpha_{ij} \right\rangle).
\end{equation}
The presence of the averaging operator is responsible for the fact that the strain energy density has a dependency on the voxel size $\Delta V$. Given a voxel size, we can now calculate both $\psi^{\rm{micro}}_{\triangle V}$ and $\psi^{\rm{meso}}_{\triangle V}$ in \eqref{energy-equation}. From this, $\triangle \psi(\bm{{x}})$ can be obtained which is the strain energy density that was lost during the coarse graining. \\
As the last step, the relation between $\deltaV$ and the continuum dislocation microstructure has to be established. It should have the form
\begin{equation}
    \deltaV=\psi(\rho,\boldsymbol{\alpha}, \ldots).
\end{equation}
There, the function arguments are density and density-like quantities, all of which dependent on the averaging voxel size. A number of different relations for energy versus dislocation densities have been formulated~\cite{zaiser2015local, Wilkens1969, berdichevsky2006continuum, groma2016dislocation} based on different assumptions. For simplicity, we firstly adopt the relation by Berdichevsky \cite{berdichevsky2006continuum}
\begin{equation} 
\psi \propto Gb^{2}\rho\rm{log}\left(\frac{\rho}{\rho_{0}}\right),
\end{equation} 
where $G$ is the shear modulus and $\rho_{0}$ is the ``screening distance''. In our analysis we take $\rho_{0}$ as $\frac{1}{a^{2}}$ where $a=1.5b$ is the dislocation core radius (alternative relations will be discussed in section~\ref{other-functionals}). Thus, we can rewrite \eqref{energy-equation} as:
\begin{equation}\label{fitting-function}
\psi^{\rm{micro}}_{\triangle V}=\psi^{\rm{meso}}_{\triangle V}+CGb^{2}\rho\rm{log}\left( \frac{\rho}{\rho_0}\right)
\end{equation}
where $C$ is a fitting parameter which is supposed to be independent of the density values $\rho$. This, however, still needs to be shown. A summary of the complete proposed strategy is as follows:

\hbox{}\hfill
\fbox{\begin{minipage}{0.8\textwidth}
For a discrete dislocation microstructure, 
\begin{enumerate}
    \item we firstly use small enough voxel size (usually $\Delta V \le b^{3}$ ) to compute;
    \item then use the desired voxel size $\Delta V$ to get $\psi^{\rm{meso}}_{\triangle V}$;
    \item compute the difference $\triangle \psi=\psi^{\rm{micro}}_{\triangle V} - \psi^{\rm{meso}}_{\triangle V}$;
    \item identify the fitting parameter $C$\footnote{The number of fitting parameters depends on the energy formulations used, in section 6, we use energy formulations that include more field variables, thus more parameters are fitted.} through \eqref{fitting-function}; 
\end{enumerate}
\end{minipage}}\hfill\hbox{}\\

\subsection{Description of the proposed strategy on 2D dislocation structure.}
\label{2D-section}
\begin{figure}[ht!] 
	\centering
	\includegraphics[scale=0.7]{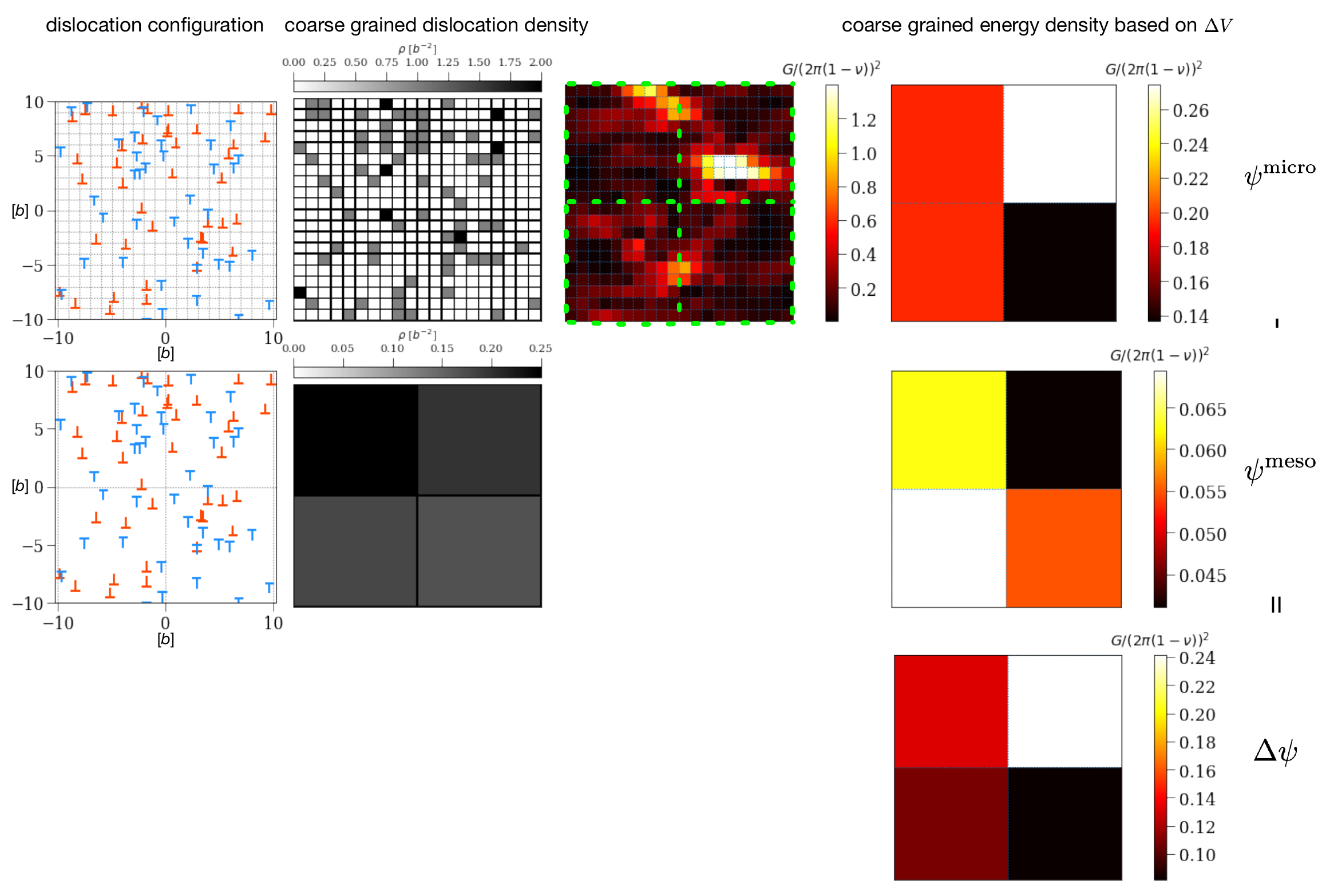}
	\caption{Schematic of the data-mining in 2D. The dislocation structure consists of different signs of dislocations (represented by different colors) is firstly coarse grained by the reference voxel of size $\Delta x=b, \Delta y=b$ as shown in the top row. With the coarse grained dislocation density, the strain energy density (the third column) can be calculated. If we average the strain energy density data in a desired coarse voxel size, for example $\Delta x=10b, \Delta y=10b$ indicated by the green dashed line, we get $\psi^{\rm{micro}}$ for the large voxel size, shown in the last column of the first row. For coarse graining voxel size $\Delta x=10b, \Delta y=10b$ as shown in the middle row, we can calculate the corresponding dislocation density and the strain energy density $\psi^{\rm{meso}}$. $\Delta \psi$, shown in the last row, is the difference between $\psi^{\rm{micro}}$ and $\psi^{\rm{meso}}$.
    }
\label{fig:2D-1}
\end{figure} 
2D DDD simulations comprising of infinite long parallel edge dislocations make it easy to obtain large amount of data and, at the same time, allows to use simplifications of the above formulations. In the following, an explicit formulations  for a two-dimensional plane strain dislocation system is derived; a typical example of a random 2D discrete dislocation system in a quadratic domain of size $L=20b$  is shown in \figref{fig:2D-1}. There, the points belonging to a coarse-graining ``voxel''\footnote{Even though ``voxel'' usually denotes a 3D object, we use it here for the 2D systems as well. This is motivated by the fact that the 2D system in fact represents a plane strain systems that is infinitely long along the third direction.} reduces to 
\begin{equation}
\Omega_{i}=
\left[x_{i}-\frac{\Delta x}{2} \ldots x_{i}+\frac{\Delta x}{2}\right]
\times 
\left[y_{i}-\frac{\Delta y}{2} \ldots y_{i}+\frac{\Delta y}{2}\right]
\end{equation}
The domain average reduces to 
\begin{equation}
\langle\bullet\rangle_{\Omega_{i}} = \frac{1}{\Delta x \Delta y}  \int\limits_{x_{i}-\frac{\Delta x}{2}}^{x_{i}+\frac{\Delta x}{2}} \int\limits_{y_{i}-\frac{\Delta y}{2}}^{y_{i}+\frac{\Delta y}{2}} \bullet \ \dx' \dy' 
\end{equation}
The dislocation density \eqref{dislocation-density} simplifies to
\begin{equation}
\rho=n_{i}/(\Delta x \Delta y),
\end{equation}
where $n_{i}$ is the number of dislocations in $\Omega_{i}$. 
In this 2D system, all dislocations have the same Burgers vector whose magnitude is $b$, their line directions are either $[0\;0\;1]$ or $[0\;0\;\bar{1}]$.
The non-zero component of the coarse grained Kr\"{o}ner-Nye tensor content contained in $\Omega_{i}$ \eqref{Nye-tensor} can be simplified to 
\begin{equation}\label{2D-Nye}
\alpha^{i}_{13} = \frac{n^{+}_{i}- n^{-}_{i}} {\Delta x \Delta y}b,
\end{equation} 
where $n^{+}_{i}$ and $n^{-}_{i}$ are the number of positive and negative dislocations in $\Omega_{i}$ (with line direction $[0\;0\;1]$ and $[0\;0\;\bar{1}]$, respectively).
The stress tensor in \eqref{stress-tensor} can then be simplified to
\begin{align}
\begin{split}
\boldsymbol{\sigma}(x_j,y_j) = \sum\limits_{i=1}^N \int\limits_{x_{i}-\frac{1}{2}\Delta x}^{x_{i}+\frac{1}{2}\Delta x}\int\limits_{y_{i}-\frac{1}{2}\Delta y}^{y_{i}+\frac{1}{2}\Delta y} \alpha^{i}_{13}\boldsymbol \sigma^\perp(x_j  - x_{i}, y_j - y_{i})\dx \, \dy \\=\sum\limits_{i=1}^N \boldsymbol \left( (n^{+}_{i} -n^{-}_{i} )\bm{\sigma}^\perp(x_j  - x_{i}, y_j - y_{i})\right)
\end{split}
\end{align}
where $\bm{\sigma}(x_j,y_j)$ is the resulting stress tensor, due to the presense of all dislocations in the system, at point ($x_{j}, y_{j}$) and represents the stress averaged in the domain $\Omega_{j}$. $\bm{\sigma}^\perp$ is the stress field of an edge dislocation with burgers vector $[b\;0\;0]$. The sum runs over all voxels in the system. For $\bm{\sigma}^\perp$, we use the non-singular solution from Cai~\cite{cai2006non} which avoids the problem of diverging stresses in the core so that the core does not have to be removed (cf. \figref{fig:3D-single-loop}, label 2.)). This stress tensor consists of the following components:
\begin{eqnarray}
\sigma_{xx}(x, y)=-\frac{G b}{2\pi (1-\nu)}\frac{y}{\rho_{a}^{2}}\left[1+\frac{2(x^{2}+a^{2})}{\rho_{a}^{2}}\right],\nonumber \\
\sigma_{yy}(x, y)=\frac{G b}{2\pi (1-\nu)}\frac{y}{\rho_{a}^{2}}\left[1-\frac{2(x^{2}+a^{2})}{\rho_{a}^{2}}\right],\nonumber \\
\sigma_{zz}(x, y)=-\frac{G b}{\pi (1-\nu)}\frac{y}{\rho_{a}^{2}}\left[1+\frac{a^{2}}{\rho_{a}^{2}}\right],\nonumber \\
\sigma_{xy}(x, y)=\frac{G b}{2\pi (1-\nu)}\frac{x}{\rho_{a}^{2}}\left[1-\frac{2y^{2}}{\rho_{a}^{2}}\right],\nonumber \\
\sigma_{xz}(x, y)=\sigma_{yz}(x, y)=0,
\end{eqnarray}
where $a$ is the dislocation core spreading radius and $\nu$ is the Poisson's ratio. Furthermore, the abbreviation $\rho_{a}=\sqrt{x^{2}+y^{2}+a^{2}}$ is used. With the average stress at the voxel center of $\Omega_{j}$, the energy density can be then calculated following \eqref{energy-density}.

For the dislocation structure shown in \figref{fig:2D-1}, a large voxel size of $\Delta x=10b, \Delta y=10b$ results in the dislocation density shown in \figref{fig:2D-1}, the second figure in the middle row; the strain energy density  $\psi^{\rm{meso}}$ is shown in \figref{fig:2D-1}, middle right. Choosing a much smaller voxel size of $\Delta x= b, \Delta y=b$, we can again compute the coarse grained strain energy density as shown in \figref{fig:2D-1}, the second figure in the top row. Averaging the strain energy density in a corresponding coarse voxel (indicated by green dashed line) would give $\psi^{\rm{micro}}$ which is shown at the top right. With $\psi^{\rm{micro}}$ and $\psi^{\rm{meso}}$, one can then get $\Delta \psi$ which is shown at the bottom right. 

Note that one has to make sure that the fine voxel size (in the top row) that is used to calculate $\psi^{\rm{micro}}$ should be small enough to accurately capture the microstructure information of the system. The convergence study shown in the appendix showed that a voxel edge length of $1b$ is small enough to accurately approximate $\psi^{\rm{micro}}$.

\section{Data analysis of 2D discrete dislocation systems}
Subsequently, the proposed data mining strategy is applied to dislocation microstructures data from 2D discrete dislocation dynamics simulations, and the data is analyzed with respect to the energy contribution that was lost during coarse graining. 

\subsection{Creation of 2D dislocation microstructures.}
\label{2D-creation}
\begin{figure}[!ht] 
	\centering
	\includegraphics[scale=0.75]{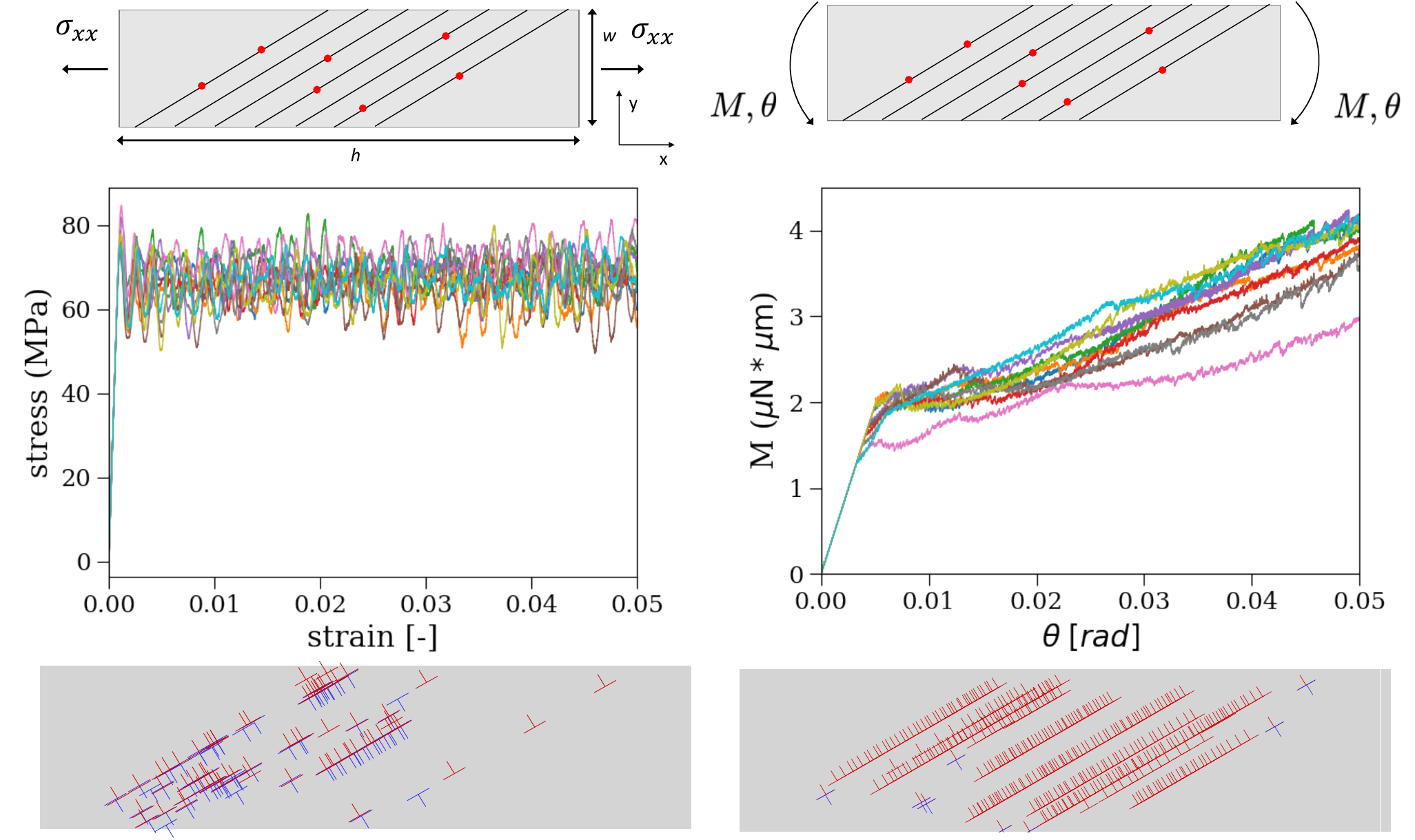}
	\caption{Top: Schematic of the two simulated problems. The red dots represent the positions of Frank-Read sources. $M$ and $\theta$ stand for the bending moment and bending angle respectively. Middle: Typical mechanical responses (10 realizations) in DDD simulations. Different realizations have the same dislocation source density, but different source positions. Left: $\sigma$ versus $\varepsilon$ in tension; Right: bending moment versus bending angle in bending; Bottom left: Typical dislocation microstructure in tension; Bottom right: Typical dislocation microstructure in bending.}
\label{fig:2D-mechanical}
\end{figure}

2D dislocation structures are obtained from previously validated and benchmarked 2D discrete dislocation dynamics simulations~\cite{papanikolaou2017obstacles}. Two different loading conditions are considered: tension where statistically stored dislocations (SSDs) dominate  and bending where geometrically necessary dislocations (GNDs) dominate. The simulated problems are shown in \figref{fig:2D-mechanical}. 
The model geometry has an aspect ratio of ${h/w}=4$ with ${w}=1\upmu\rm{m}$. Both simulation types are carried out by strain control (constant $\dot{\varepsilon}$ in tension and constant $\dot{\theta}$ in bending). A single slip system is used with a slip plane spacing of $10b$. The slip system is oriented at $30^{\circ}$ relative to the $x$ direction. The red dots represent the locations of Frank-Read (FR) dislocation sources. The strength of these sources follows a Gaussian distribution with an average value of 50\,MPa and standard deviation of 10\,MPa. The sources will nucleate a dislocation dipole once the resolved shear stress on the source site reaches the source strength for a time period of $t_{\rm{nuc}}$, this process mimics the bow out of FR dislocations in 3D.  
 Further details of the 2D DDD simulations can be found in~\cite{papanikolaou2017obstacles, cleveringa1999discrete}. 

Typical mechanical responses are shown in \figref{fig:2D-mechanical} (middle row) for multiple realizations where the dislocation source position differs while keeping the total source density the same. It can be seen in \figref{fig:2D-mechanical} that in tension, the stress-strain curves show almost elasto-perfectly plastic behaviors. By contrast, in bending, hardening appears after the initial yielding, and the reason is due to plastic strain gradient and the formation of GNDs~\cite{kubin2003geometrically, gao2003geometrically, fleck2003role, acharya2003geometrically}. Typical dislocation structures for the two different loading systems can be seen in \figref{fig:2D-mechanical} (bottom row) where it can be seen that in tension mainly dislocation dipoles/SSDs occur while GNDs dominate in bending. 

\subsection{Retrieving the ``lost'' energy for 2D dislocation structures.}
\label{2DDDD-analysis}
\begin{figure}[!hb] 
	\centering
	\includegraphics[width=1.0\textwidth]{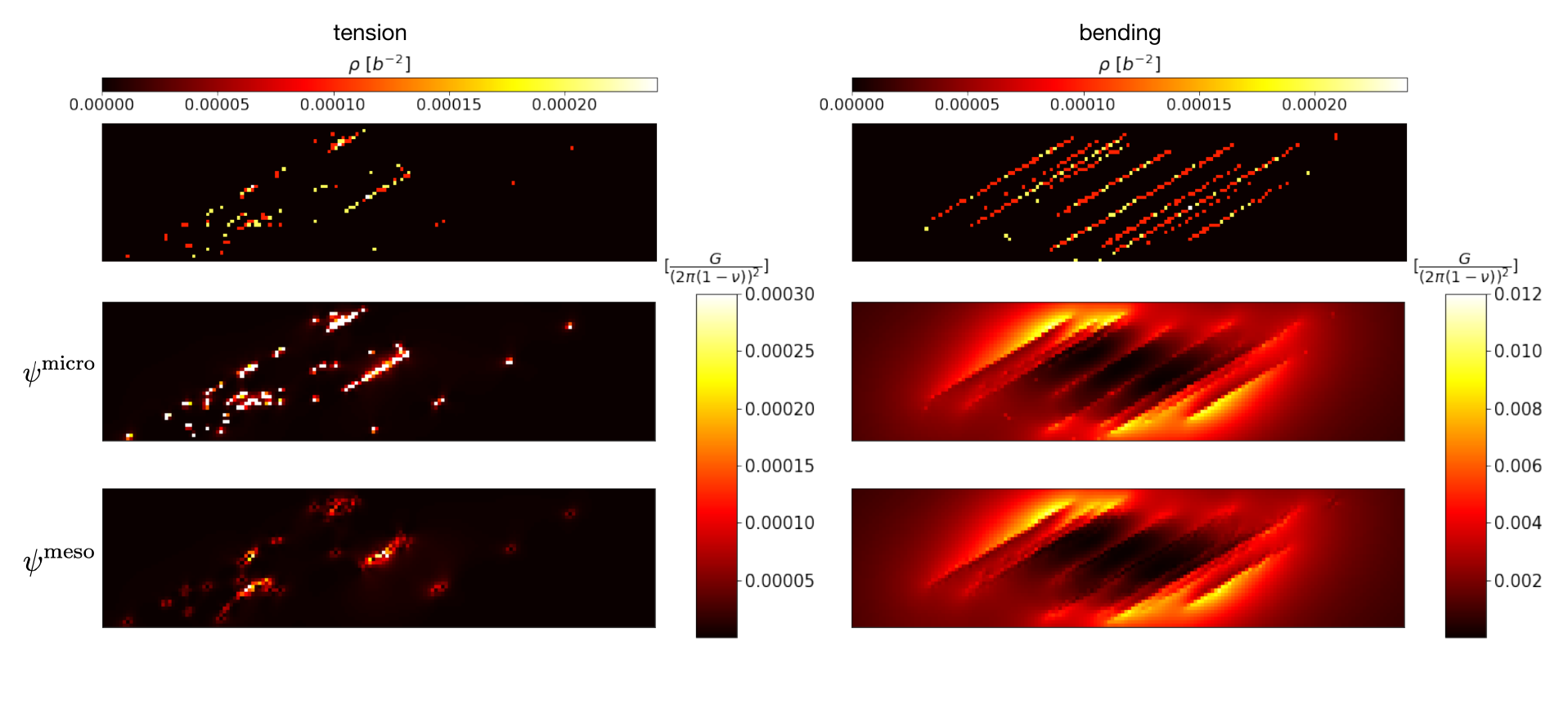}
	\caption{Top row: coarse grained dislocation density. Middle left: $\psi^{\rm{micro}}$ in tension; Bottom left: $\psi^{\rm{meso}}$ in tension; Middle right: $\psi^{\rm{micro}}$ in bending; Bottom right: $\psi^{\rm{meso}}$ in bending.
}
\label{fig:2D-energy-density}
\end{figure}

\begin{figure}[!h] 
	\centering
	\includegraphics[scale=0.8]{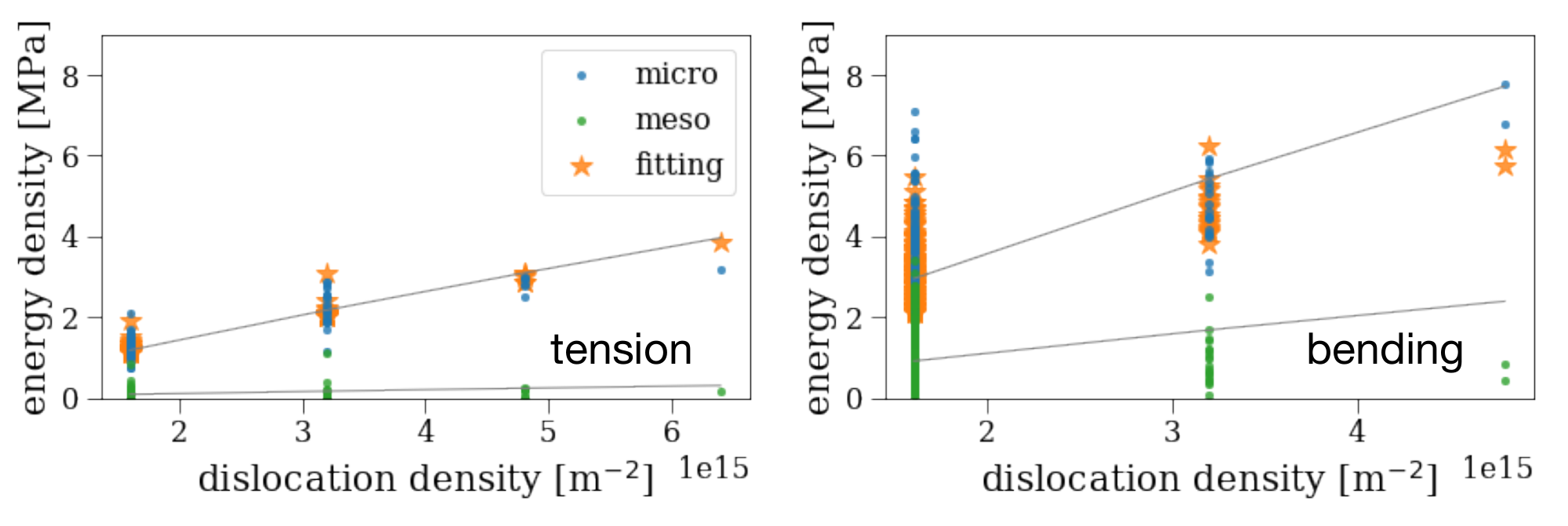}
	\caption{Strain energy density as a function of dislocation density in the voxel of $\Delta x = \Delta y=100b$. Analysis of the energy based on a dislocation configuration obtained through (left) tension and (right) bending. The energy function $Gb^{2}\rho\rm{log}\left(\frac{\rho}{\rho_{0}}\right)$ is plotted as solid lines.  $\psi^{\rm{meso}}_{\triangle V}+CGb^{2}\rho\rm{log}\left(\frac{\rho}{\rho_{0}}\right)$ is denoted by asterisks in black.}
\label{fig:2D-energy-fit}
\end{figure}

For a given dislocation structure and voxel size, following the method explained in the previous section, we can get $\psi^{\rm{micro}}_{\triangle V}$ and $\psi^{\rm{meso}}_{\triangle V}$. One example is shown in \figref{fig:2D-energy-density} based on the dislocation structures in \figref{fig:2D-mechanical} with the voxel size taken as $\Delta x = \Delta y=100b$.  
Coarse grained dislocation density is shown in \figref{fig:2D-energy-density} top row, through coarse grained dislocation density, we can then get coarse grained energy density $\psi^{\rm{meso}}$ (bottom row). Using reference voxel size $\Delta x = \Delta y=b$ then average energy density in a coarse voxel size of $\Delta x = \Delta y=100b$, we obtain $\psi^{\rm{micro}}$ (bottom row). It can be seen that for tension, there is clear difference between $\psi^{\rm{micro}}$ and $\psi^{\rm{meso}}$ while the difference is quite small for bending. The difference can be better viewed in \figref{fig:2D-energy-fit} where 
we plot the energy density data as a function of the coarse grained dislocation density. First of all, it can be seen that the 
chosen function for the energy density (plotted as solid lines)  $Gb^{2}\rho\,\rm{log}\left({\rho}/{\rho_{0}}\right)$ has good agreement with the numerical data labeled ``meso''. Therefore, it is expected that $\psi^{\rm{meso}}_{\triangle V}+Gb^{2}\rho\,\rm{log}\left({\rho}/{\rho_{0}}\right)$ (plotted as asterisks in black) has a good fit with $\psi^{\rm{micro}}_{\triangle V}$. 
Secondly, it can be noticed that, compared to the bending data, $\psi^{\rm{meso}}_{\triangle V}$ in case of tension is much smaller than the corresponding $\psi^{\rm{micro}}_{\triangle V}$, i.e., a significant amount of energy density related to the microstructure has been lost. This is because the dislocation microstructure mainly consists of SSDs (dislocation dipoles), therefore the coarse graining process can result in a complete loss of the dipole information due to the fact that the dipole has no contribution to $\alpha^{i}_{13}$ in \eqref{2D-Nye}, i.e., no long range stress, therefore no contribution to the strain energy density. Furthermore, we can also see that for bending where GNDs dominate, the values of  $\psi^{\rm{meso}}_{\triangle V}$ are no longer as small as in tension. The difference between $\psi^{\rm{meso}}_{\triangle V}$ and $\psi^{\rm{micro}}_{\triangle V}$ mainly results from the loss of the dislocation spatial information in the coarse grained voxel since $\alpha^{i}_{13}$ has no information of the exact location of dislocations inside the region $\Omega_{i}$. 

\begin{figure}[!h] 
	\centering
	\includegraphics[scale=0.8]{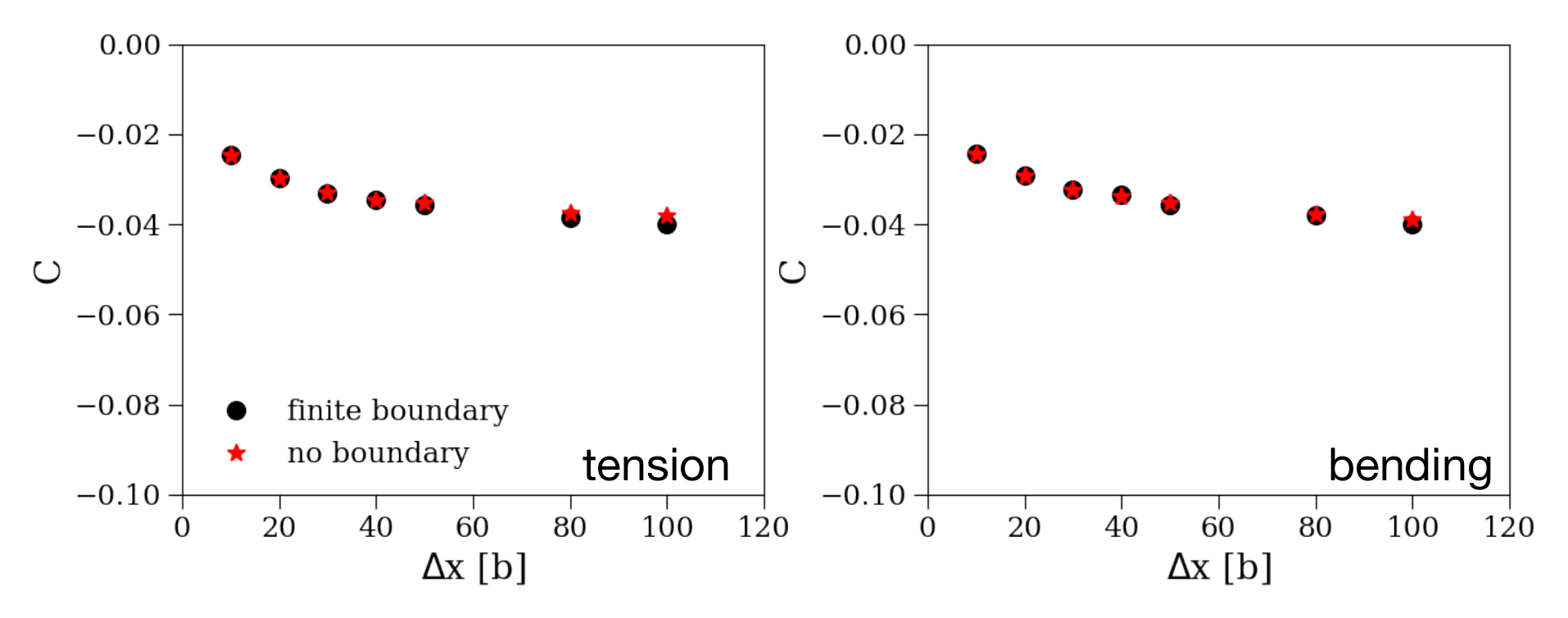}
	\caption{Dependence of the fitting parameter $C$ on the voxel size, (left) tension and (right) bending. The red asterisks are data without consideration of the finite boundary while the black dots are data that include the finite boundary effect.}
\label{fig:2D-fit-voxelsize}
\end{figure}

Clearly, this process is ``mesh dependent'' -- a fact that can not be avoided. Therefore, the influence of different coarse graining voxel sizes were tested together with the above process. 
The fitting parameter $C$ (in equation~\ref{fitting-function}) as the function of the voxel size is shown in \figref{fig:2D-fit-voxelsize} as red stars: $C$ is seen to converge to the value around -0.04 for larger voxel sizes.

However, in the above analysis, we utilized the analytical solution of dislocations in an infinite domain while our sample geometry domain is finite, therefore the stress field of dislocations needs to be corrected due to the free surfaces. The way we correct the boundary effect is through the Van der Giessen-Needleman superposition~\cite{vandergiessen1995} that utilizes finite element method (FEM). For both loading conditions (tension and bending), the actual boundary for the dislocation structure is chosen as the following: for the loading edges, i.e., left and right edges as shown in \figref{fig:2D-mechanical}, the displacement boundary condition $u_{y}=0, u_{x}(x=\frac{h}{2})=0$ is used. This boundary condition corresponds to the initial status of the loading process, thus the strain energy density $\psi^{\rm{ext}}$ which is related to the external loading is separated. Since the superposition method is FEM-based, note that there FEM mesh size is larger than the coarse graining voxel size, therefore the stress field is interpolated to the voxel centers to get the stress under the given boundary conditions. {
Detailed information can be found in the Appendix. The fitting parameters $C$ for systems with finite boundaries  is also shown in \figref{fig:2D-fit-voxelsize}. It can be seen that the effect of the finite boundary on the fitting parameter $C$ is quite limited, therefore for the 3D dislocation structures, we also ignore the effect of the finite boundary for the simplicity of the calculation even though such approximation should be handled with care for very  small geometry where the surface effect becomes stronger.

\section{Data analysis of 3D discrete dislocation systems}

\begin{figure}[!ht] 
	\centering
	\includegraphics[width=0.9\textwidth]{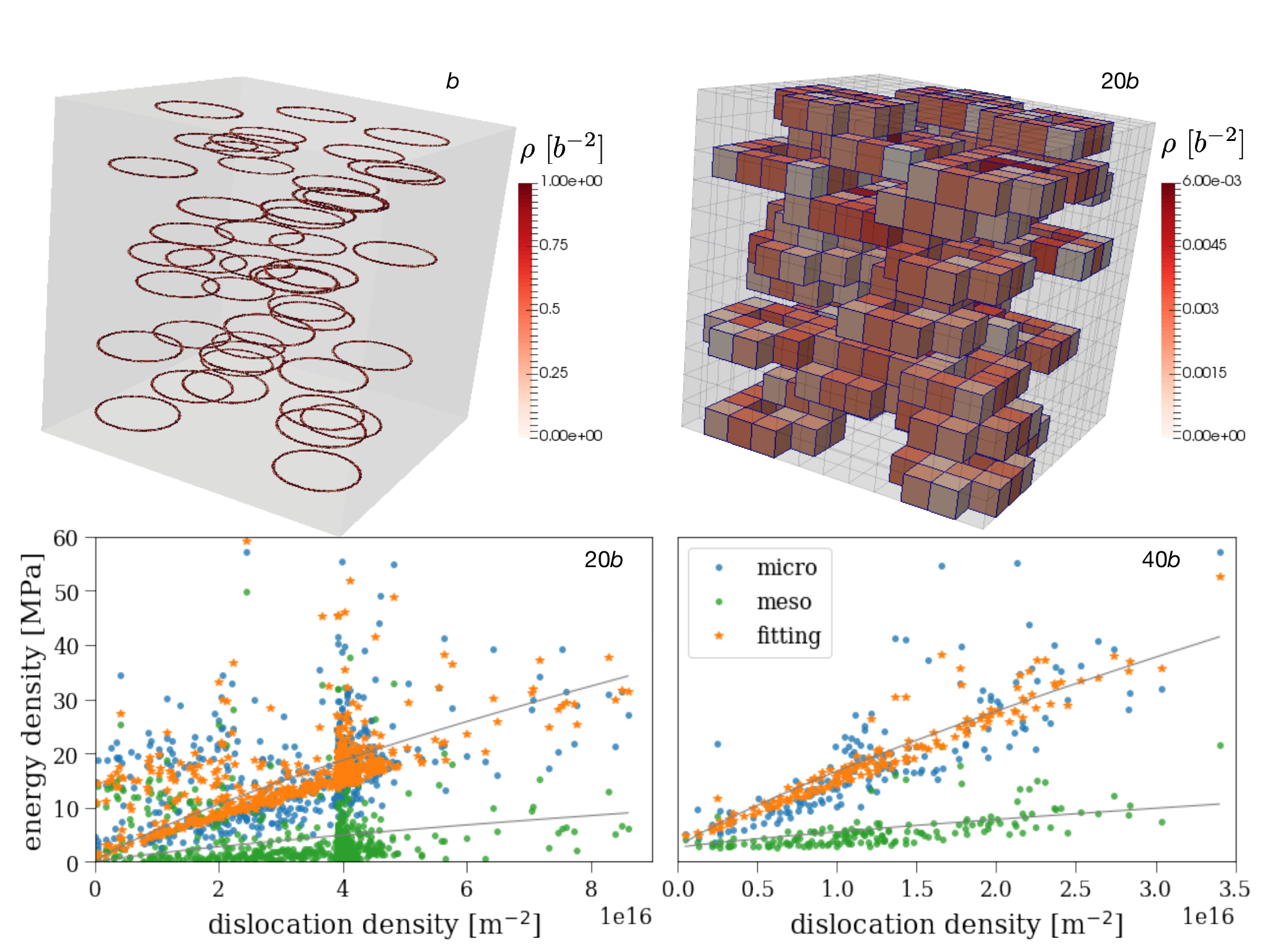}
	\caption{Top left: dislocation density field of random dislocation loops when the voxel size is $b$; Top right: coarse-grained continuum dislocation density field when the voxel size is chosen as 20$b$; Bottom left: analysis of the energy density when the coarse voxel size is chosen as 20$b$; Bottom right: analysis of the energy density when the coarse voxel size is 40$b$. The domain size is 60nm. The energy function $Gb^{2}\rho\rm{log}\left(\frac{\rho}{\rho_{0}}\right)$ is plotted as solid lines.  $\psi^{\rm{meso}}_{\triangle V}+Gb^{2}\rho\rm{log}\left(\frac{\rho}{\rho_{0}}\right)$ is denoted by asterisks in orange. The corresponding voxel size in each data plot is on the top right of each figure.}
\label{fig:3D-loop}
\end{figure}

In the previous section, we have applied our data-mining method to recover missing energy density contributions in the context of 2D continuum dislocation density fields obtained through discrete dislocation dynamics simulations. In this section, we now apply the proposed method to two different 3D dislocation microstructures: random distributions of dislocation loops and dislocation microstructures obtained from a relaxation of 3D DDD using the code MODEL~\cite{po2015mechanics}. Due to the high computational cost in the 3D analysis, we chose two relative small sample sizes: 30nm and 60nm. Results from such small sample sizes may be somewhat limited in terms of predictions for larger systems. However, the 2D study above showed that free surfaces do not have a significant influence on the parameterisation of the energy density. Furthermore, the advantage is that voxel sizes as small as $b^{3}$ can still computationally be handled. 

The random loop structure is shown in \figref{fig:3D-loop} which is the coarse grained density field using a voxel size of $b$. Dislocation loops have the radius of $R=30b$, they are placed parallel to the $x-y$ plane in a cubic domain of above mentioned sizes.  The total number of loops is chosen such that the dislocation density is 2$\times 10^{16}\rm{m}^{-2}$. The resulting dislocation density for a coarse graining voxel size of $20b$ can be seen in \figref{fig:3D-loop}(top right). There, voxels with density values very close to or exactly zero are made translucent. In \figref{fig:3D-loop} bottom we show the analysis of the energy density when the coarse voxel size is $20b$ and $40b$, respectively.  It is seen that the proposed method still works reasonably well. Note that in \figref{fig:3D-loop}(bottom left) when the voxel size is 20$b$ (we observe the same phenomenon when the voxel size is even smaller), there is a clustering of the data around certain density value. This is caused by the synthetic loop configuration: when the voxel size is small enough, a loop can pass through many voxels, yet each voxel contains similar arc length which results in a similar dislocation density. 

\begin{figure}[!h] 
	\centering
	\includegraphics[scale=0.8]{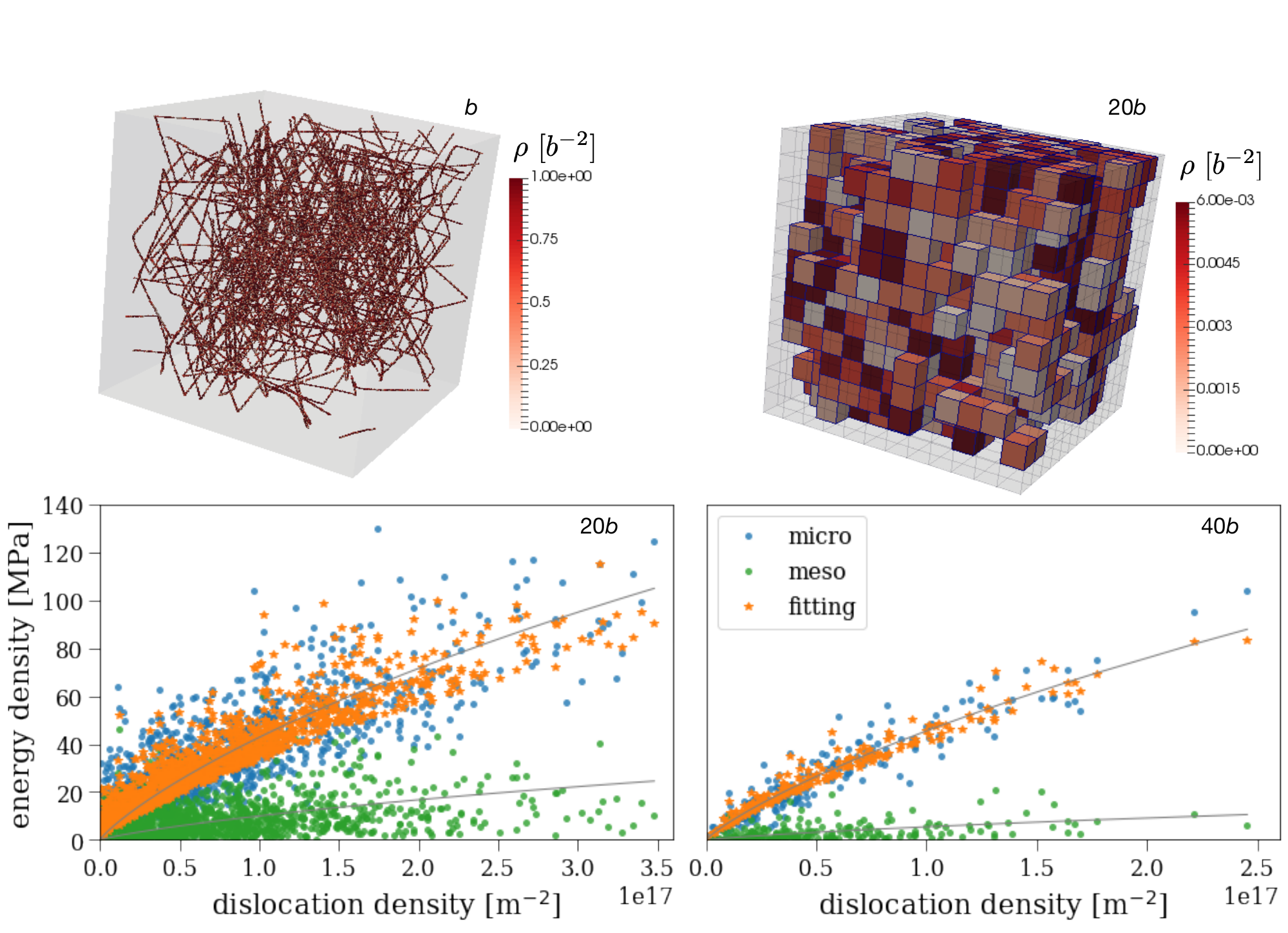}
	\caption{Top left: dislocation density field of relaxed dislocation microstructure when the voxel size is $b$. The field approaches a discrete description of the dislocation microstructure; Top right: coarse-grained continuum dislocation density field when the voxel size is chose as 20$b$; Bottom left: analysis of the energy density when the coarse voxel size is chosen as 20$b$; Bottom right: analysis of the energy density when the coarse voxel size is 40$b$. The energy function $Gb^{2}\rho\rm{log}\left(\frac{\rho}{\rho_{0}}\right)$ is plotted as solid lines.  $\psi^{\rm{meso}}_{\triangle V}+Gb^{2}\rho\rm{log}\left(\frac{\rho}{\rho_{0}}\right)$ is denoted by asterisks in orange. The corresponding voxel size in each data plot is on the top right of each figure.}
\label{fig:3D-realistic}
\end{figure}
Furthermore, we also analyze the relaxed dislocation microstructures (whose initial configuration consists of dipolar loops) from 3D DDD simulations. These structures also have been recently analyzed in a machine learning based  study~\cite{steinberger2019machine} for understanding the importance of different microstructural features (density fields in continuum dislocation theory). Similar to \figref{fig:3D-loop} top row, we show the coarse grained dislocation density field of different voxel sizes in \figref{fig:3D-realistic}. At the same time, one can also notice that in a realistic dislocation microstructure, less dislocations exist in the region close to the surface due to the image force at the sample free surface. As shown in \figref{fig:3D-realistic} bottom row for different voxel sizes, the proposed method also works reasonably well for the realistic dislocation microstructures. 

\begin{figure}[!h] 
	\centering
	\includegraphics[scale=0.6]{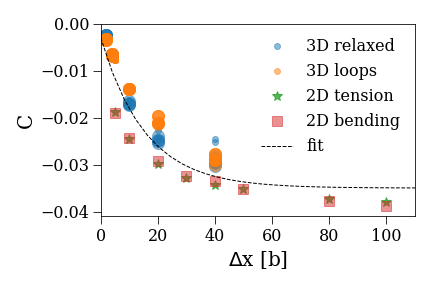}
	\caption{Summary of the fitting parameters for 2D and 3D dislocation microstructures. For 2D results, each data point is the average of 30 samples. For 3D results, the symbol size represents the sample size since both 30nm and 60nm sample are analyzed.}
\label{fig:data-summary}
\end{figure}
The summary of the fitting parameters for 2D (30 tension and 30 bending samples) and 3D dislocation microstructures (12 structures each for random loops and relaxed microstructures) is shown in \figref{fig:data-summary}. It is clearly visible 
that the fitting parameter approaches zero when the voxel size approaches $b$, i.e., no information is lost when the voxel size is small enough. The fitting parameter decreases when the voxel size increases for both random loop structure and relaxed structure. At the current stage, it is computationally difficult to analyze a large enough 3D dislocation structure using many large voxels to get good statistics. Thus the analyzed 3D data is still strongly changing as the voxel size changes. However, the fitting parameter for the 3D systems follows the same trend as the parameter for the 2D system. Furthermore, even the value of the 3D parameter is quite comparable roughly following $C^{\rm 3D}\approx 0.65 C^{\rm 2D}$ where values for straight dislocations tend to be more similar to the 2D values. Overall, also for the 3D systems the strong flattening behavior as for the 2D system is expected. 

In conclusion, after assuming a simple functional form for the energy density we have now identified the relevant parameter together with the dependency on the coarse graining scale $\triangle x$ in the following form:
\begin{align}
    \psi^{\rm{micro}}_{\triangle V} &= \psi^{\rm{meso}}_{\triangle V}+CGb^{2}\rho\,\rm{log}\left( \frac{\rho}{\rho_0}\right)\\\nonumber
    \text{with} \qquad C &\approx 0.032  \exp(- 0.064  \triangle x) -  0.035
\end{align}
Assuming an exponential function, the particular form of the fit parameter $C$ as function of the coarse graining size is shown in \figref{fig:data-summary} (dashed line). With these information now a CDD simulation could be performed.

\section{Further discussions on other forms of defect energy functional}
\label{other-functionals}
So far, the strain energy density was assumed to follow the functional form of $Gb^{2}\rho\rm{log}\left(\frac{\rho}{\rho_{0}}\right)$, which implies that the energy density only depends on the dislocation density. Nonetheless, one also has other choices in terms of the physical interpretation of the energy contribution. For example,  for the quasi 2D case, with the low GND approximation, we can also write $\triangle \psi$ following Groma et al.~\cite{groma2007dynamics} :
\begin{equation}
\label{function}
\triangle \psi=AGb^{2}\rho\rm{log}\left(\frac{\rho}{\rho_{0}}\right)+\it{B} \mathrm{\frac{\gndrho^2}{2\rho}}.
\end{equation}
\begin{figure}[!h] 
	\centering
	\includegraphics[scale=0.5]{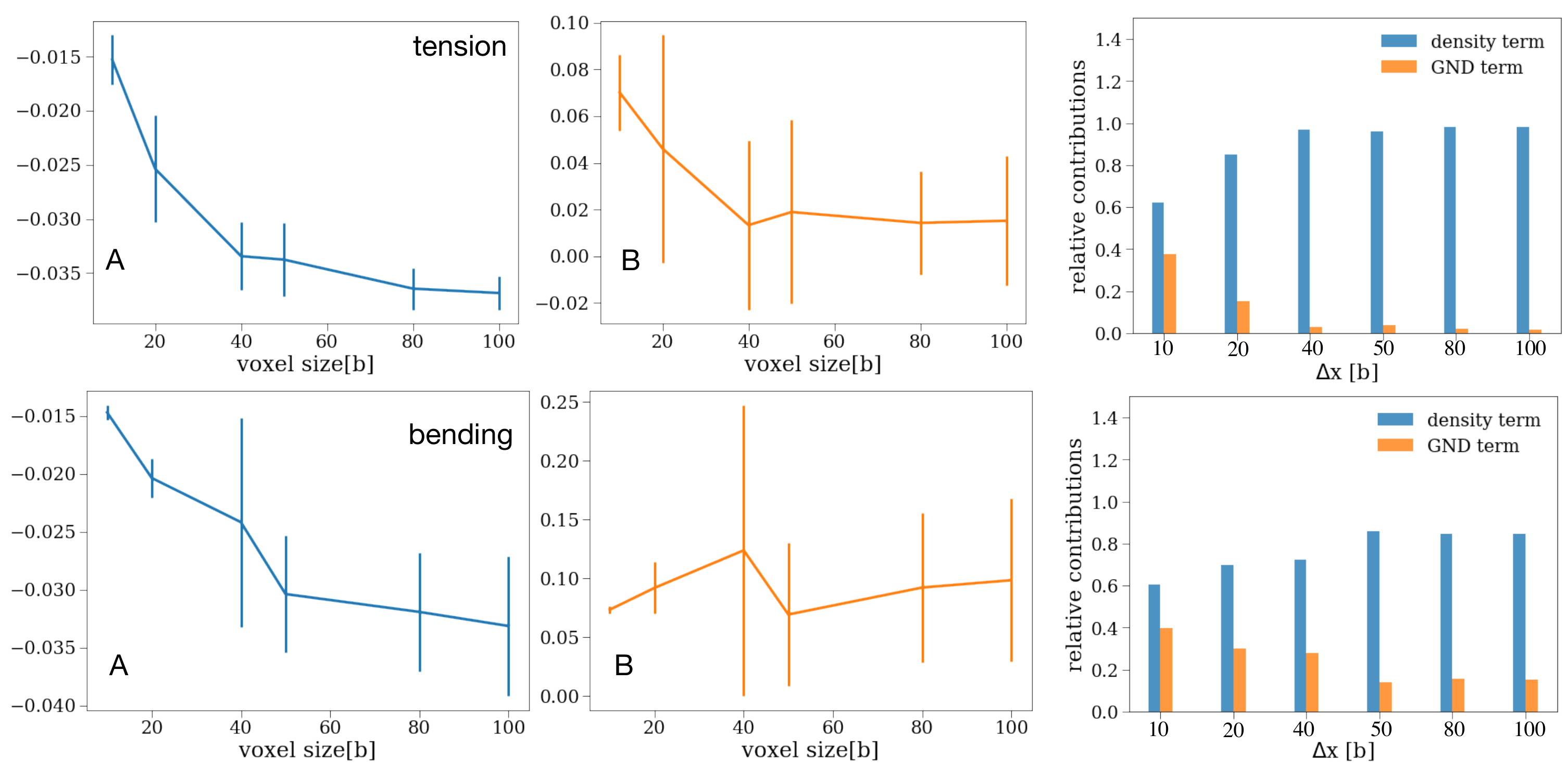}
	\caption{Fitting parameters based on dislocation microstructures obtained in tension (top row) and bending (bottom row). The error bar is the standard deviation of 20 realizations. In the right column, the contribution to the energy from each term, are shown for different voxel sizes. }
\label{fig:two-parameters-fit}
\end{figure}
In the above formulation, the defect energy density does not only depend on the total dislocation density, but also on the GND density $\gndrho$. The parameters $A$ and $B$ quantifies the contribution of the two terms, and can be numerically determined, e.g. through a non-linear least square fitting method, based on our 2D DDD data. The formulation was proposed assuming a low GND density as compared to the total density.

We determine the parameters $A$ and $B$ based on typical dislocation microstructures in tension (low GND density) and bending (high GND density, thus partially violating the underlying assumption for this formulation). The results are summarized in \figref{fig:two-parameters-fit} where the first two figures on the top row are the two fitting parameters based on the low GND density configurations while the two on the bottom row are based on high GND density configurations. Not surprising, it is found that for the low GND configurations the value of the fitting parameter $A$ is very similar to the values shown in \figref{fig:2D-fit-voxelsize} where the contribution of GND is ignored. 
It is also seen that the fitting parameter $B$ (the GND term) actually converges to a value around 0.01 when the voxel size increases and the value of $A$ converges to -0.04. By contrast, the fitting parameters based on the high GND density configuration show different features: the absolute value of fitting parameter $A$ is smaller than it is for the low GND configurations since the GND term has a non-negligible contribution. The identified value of parameter $B$ roughly takes a constant value of $\approx 0.1$, i.e., the fitting parameter of the GND term does not exhibit clear dependence on the voxel size. This is not surprising since for a extreme GND configuration where GNDs are uniformly distributed in the volume, the GND density $\gndrho$ should not have  any dependence on the voxel size at all. 
Furthermore, we can also check the contribution of each term to the total strain energy density, shown on the right in \figref{fig:two-parameters-fit}. It can be seen that for dislocation microstructure created through tension, the energy contribution of the GND term decreases with increasing voxel sizes: when the voxel size is larger than 40$b$, the contribution is negligible compared to the dislocation density term. However, in the case of bending, the GND term still has more than 10\% contribution to the overall energy. This suggests the necessity of the GND term to the strain energy density calculation in GND-rich dislocation microstructures created by loading conditions such as bending, torsion and indentation.

For three dimensional curved dislocations, Hochrainer~\cite{hochrainer2016thermodynamically} proposed a thermodynamically consistent energy functional, 
\begin{equation}
\label{three-term-function}
\psi=Gb^{2}\left( A\rho\rm{log}\left(\frac{\rho}{\rho_{0}}\right)+ \frac{1}{2} \frac{\mathit{B_{ij}} \gndrho_{i}\gndrho_{j}}{\rho}+ \mathit{C}\frac{1}{2}\frac{\mathit{q}^2}{\rho^{2}} \right),
\end{equation}
with dimensionless constants $A$, $B_{ij}$, $C$. Furthermore, in a spatially and elastically isotropic theory, the tensor $\mathbf{B}$ is expected to be spherical, i.e $B_{ij}=B\delta_{ij}$ with $B$ being a constant.  The fitting parameters are shown in \figref{fig:60nm-loops} for dislocation loop microstructures. It can be seen that, compared to the single parameter fitting, the fitting parameter $A$ of the dislocation density term has similar values. Moreover, it is observed that the fitting parameter $B$ does not have clear dependence on the voxel size and $C$ becomes almost a constant when the voxel size is larger than 10$b$ (considering the large fluctuations in the values). \figref{fig:60nm-loops}(bottom right) shows the energy contribution of each term (in Eq.~\ref{three-term-function}) to the total strain energy density. It is seen that the GND contribution only dominates when the voxel size is extremely small  (2$b$). The contribution of the curvature density term increases with increasing voxel sizes. This result on the one hand shows the nature/problem of the coarse graining, i.e., the energy information related to microstructure details, such as dislocation curvature will be completely lost if one  considers only the coarse grained total dislocation density; on the other hand this also emphasizes the importance of incorporating the contribution of the curvature term in a energy density functional for 3D continuum dislocation structures.  
\begin{figure}[!ht] 
	\centering
	\includegraphics[scale=0.7]{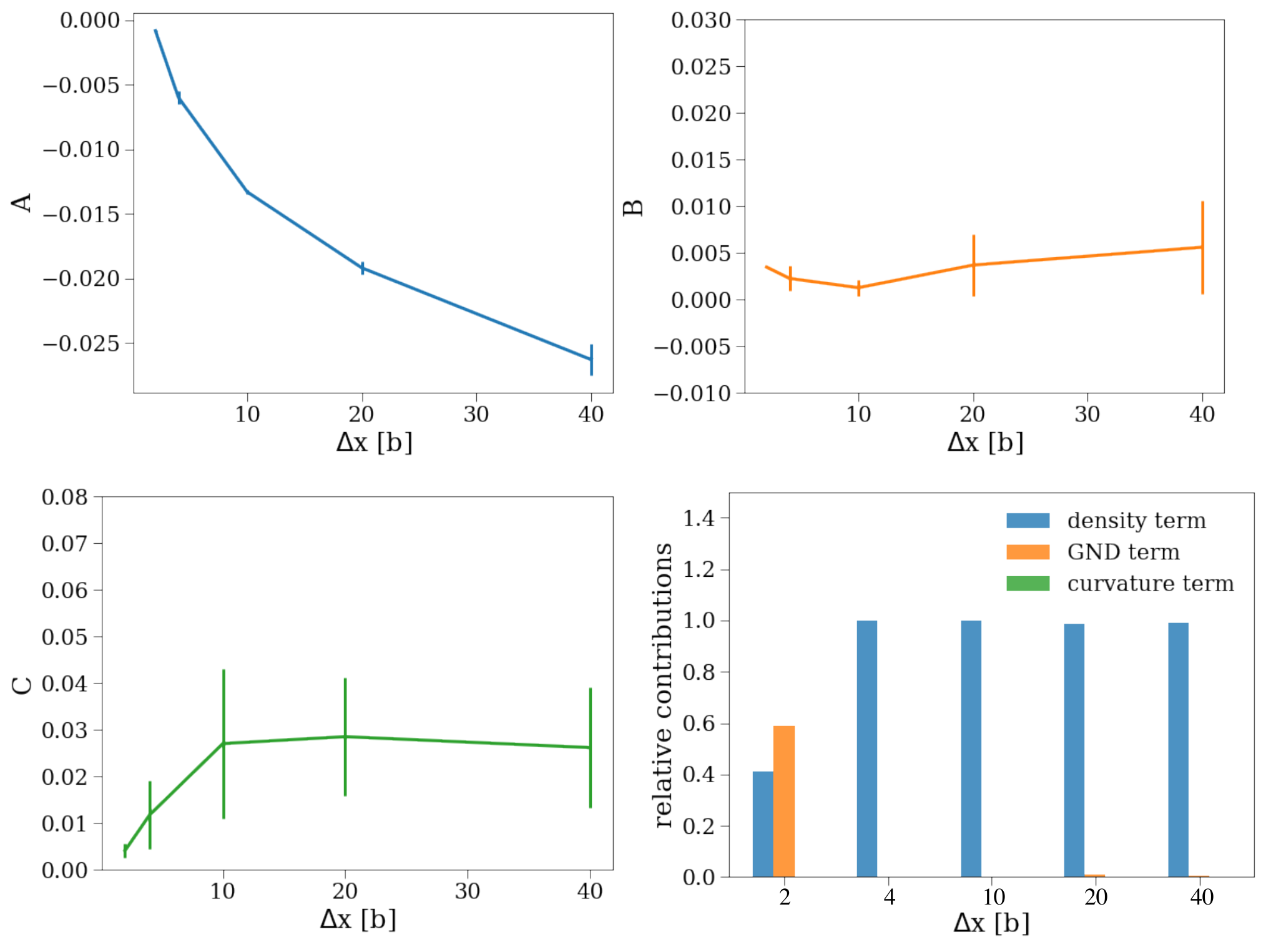}
	\caption{Fitting parameters based on dislocation microstructures of random loops. Fitting parameters as a function of the voxel size are shown. The error bar is the standard deviation of 10 realizations. Curve colors correspond to the field quantity in the bottom right figure where the energy contribution of each term to the total strain energy density is shown.}
\label{fig:60nm-loops}
\end{figure}

Finally, the fitting parameters are also calculated based on relaxed dislocation microstructures, shown in \figref{fig:60nm-relaxed}. The results show similar values for $A$ (dislocation density term) and $B$ (GND term) . However, the contribution of the curvature density term is very small (\figref{fig:60nm-relaxed}(bottom left)). The {contribution of each term} in \figref{fig:60nm-relaxed}(bottom right) shows that the GND term dominates only when the voxel size is as small as 2$b$. For larger voxel sizes, the GND contribution and curvature density contribution are essentially negligible compared to the dislocation density contribution. This indicates that for dislocation microstructures mainly containing straight segments, it is feasible to only consider the total density for the energy density calculation.

\begin{figure}[!ht] 
	\centering
	\includegraphics[scale=0.7]{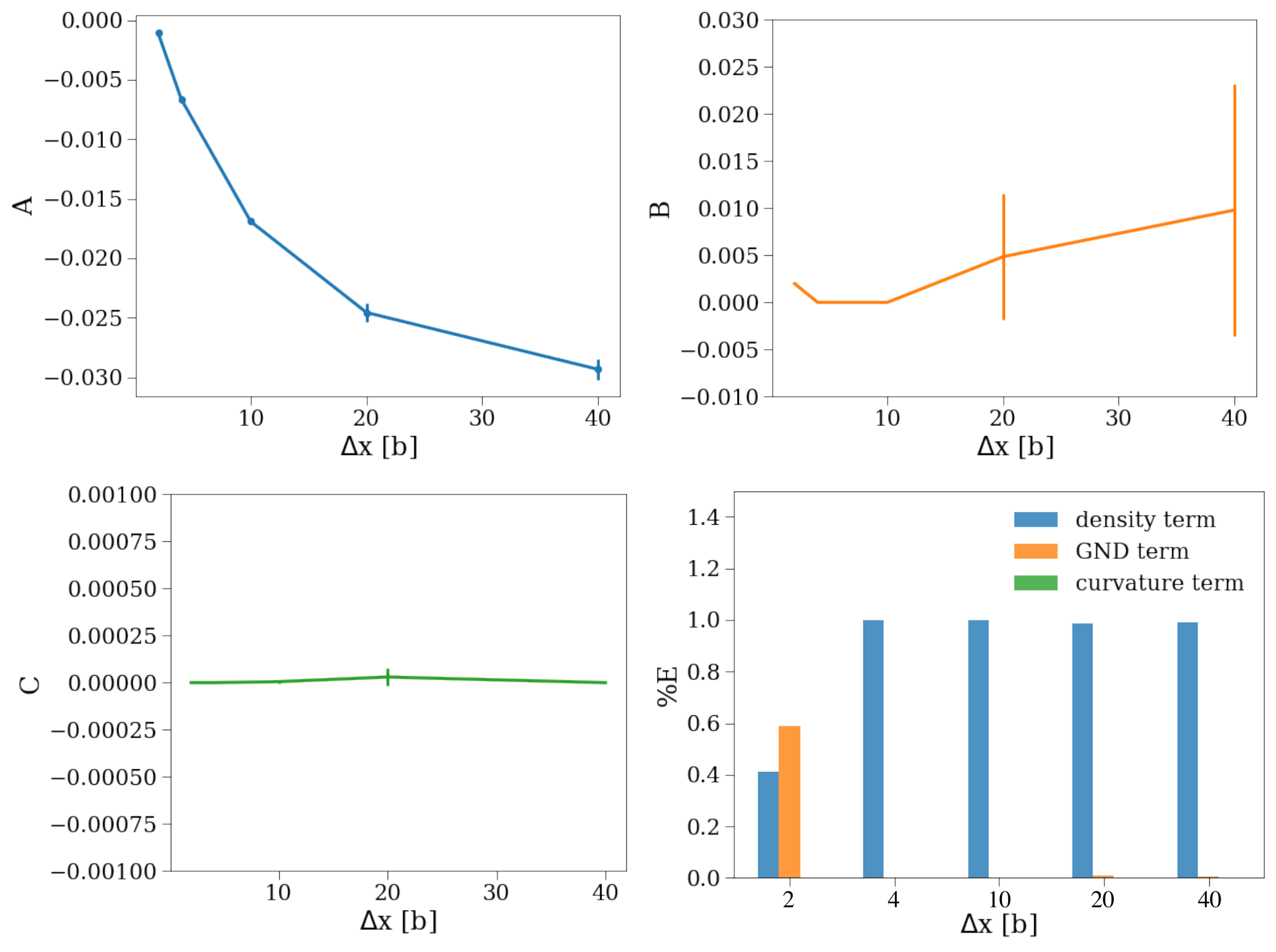}
	\caption{Fitting parameters based on relaxed dislocation microstructures. Fitting parameters as a function of the voxel size are shown. The error bar is the standard deviation of 5 realizations. Curve colors correspond to the field quantity in the bottom right figure where the energy contribution of each term to the total strain energy density is shown. }
\label{fig:60nm-relaxed}
\end{figure}

\section{Summary and conclusions}
We formulated a data-mining strategy to express the strain energy density of a dislocation system as a function of dislocation density field variables and as a function of the coarse graining voxel sizes. This became possible through the consistent interpretation of averaged quantities on different scales, and in particular through the consistent consideration of suitable reference values. The theoretical formulation of the data-mining strategy is a prerequisite for using data from DDD simulations to solve the dynamic closure problem in continuum dislocation dynamics theories. 
This problem results from that fact that unlike in discrete description of dislocation systems, the calculation of strain energy density is not trivial in a continuum dislocation framework: during the coarse graining process structural information and details about the position of dislocations is lost which, as we showed, may have a huge influence on the energy density. 
For our benchmark problems we considered a number of commonly used functional forms for the energy density as a function of continuum variables. Both 2D and 3D dislocation microstructures were used to calculate the `exact' energy density by using very fine discretizations, i.e. voxels with edge length of around one Burgers vector size. Then, the calculation of the energy density was repeated by using larger voxel sizes that are more feasible for continuum models. Based on this one can then determine parameters that occur in assumed energy functionals. 
%
%
Our numerical results shows a systematic dependency on the averaging voxel size that we quantified and which coincided for the 2D and 3D and studies. 
We then conducted our investigations for three  different  energy density formulations, including one that considers the curvature and/or an additional GND contribution. Our study revealed that these two terms are negligible if the voxel size is considerably large, as it usually always is in the case of continuum models. 
   Our numerical results suggest that depending on the type of dislocation microstructures, one can choose the proper energy formula from the computational perspective: for dislocation loop structures, both the dislocation density and curvature density term need to be considered while for straight dislocations dominated microstructures, one can only consider the energy contribution of the total dislocation density.  

This data-mining strategy is the first step towards fulfilling the \textit{dynamic consistency} of the CDD theory. The direct comparison of CDD simulation result built upon the results of this work with the corresponding DDD result will be an interesting part of our future work and thereby will allow to perform continuum simulations of dislocations dynamics in full agreement with DDD simulations but with the benefit of a significantly better computational performance.

\section{Acknowledgments}
H. Song and S. Sandfeld acknowledge financial support from the European Research Council through the ERC Grant Agreement No. 759419 MuDiLingo (“A Multiscale Dislocation Language for Data-Driven Materials Science”). 


\setcounter{figure}{0}    
\renewcommand{\thefigure}{A.\arabic{figure}} 
\section{Appendices}
\subsection{Consideration of the boundary during energy density calculation}
\begin{figure}[!ht] 
	\centering
	\includegraphics[scale=0.6]{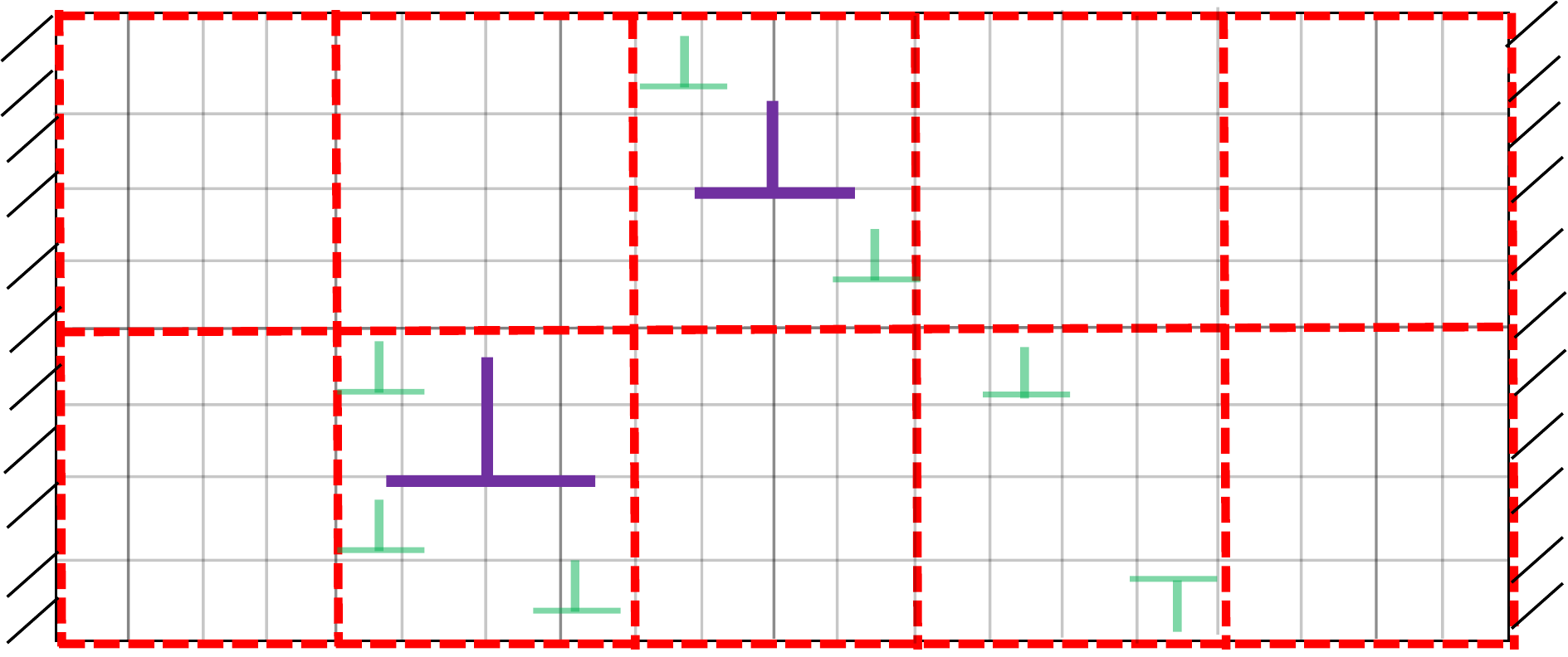}
	\caption{Schematic of boundary value problem of coarse grained dislocation configurations.}
\label{fig:boundary-problem}
\end{figure}
\figref{fig:boundary-problem} shows the schematic of how the boundary effect is considered during the energy density calculation for the coarse grained dislocation microstructure. The black lines indicates the FEM meshes while the dashed red line stands for coarse graining voxels. During the coarse graining, the dislocations (green symbols) are coarse grained within each voxel, i.e. the equivalent dislocation (which is calculated by the dislocation excess density multiplied by the voxel volume) lies in the voxel center. For the coarse grained dislocation microstructure (purple dislocations), we can easily consider the boundary effect through Van der Giessen-Needleman superposition method~\cite{vandergiessen1995}. 

\subsection{Effect of the reference voxel size}
\begin{figure}[!h] 
	\centering
	\includegraphics[scale=0.5]{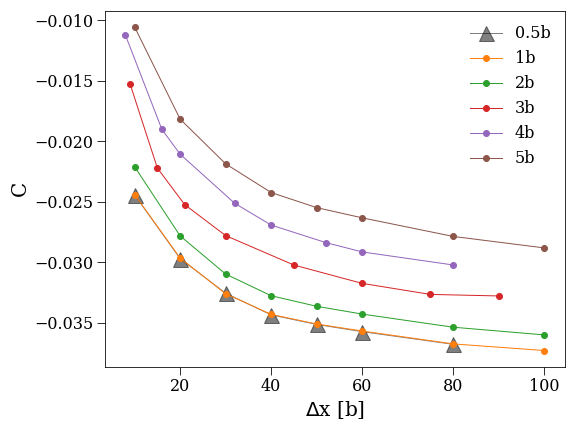}
	\caption{Fitting parameters as a function of the reference voxel size.}
\label{fig:reference-size}
\end{figure}
As we mentioned in the main text that we use voxel size of $b$ as the reference voxel size, i.e., this size is small enough to accurately calculate $\psi^{\rm{micro}}$ when using non-singular dislocation theory. In order to verify this, we have recalculated results in \figref{fig:2D-fit-voxelsize} with different reference voxel size, the results are summarized in \figref{fig:reference-size}. It can be seen that results converge when the reference voxel size is $b$. This can also be understood from fact that since we use the nonsingular analytical solution of the dislocations where we have chosen the dislocation core radius to be 1.5$b$, therefore, it is expected for the strain energy density calculation, a voxel size that is smaller than the core radius will converge. Furthermore, such convergence does not exist if one uses the singular solution of dislocations. 
\section*{References}
\bibliographystyle{ieeetr}
\bibliography{D2C}
\end{document}